\documentclass[11pt,a4paper]{article}
\pdfoutput=1
\usepackage{jheppub}
\usepackage{rotating}
\usepackage{array}
\usepackage{amsmath}
\usepackage[normalem]{ulem}
\usepackage{slashed}
\usepackage{booktabs}
\usepackage[pdftex,table]{xcolor}
\usepackage{siunitx}
\usepackage{enumerate}
\usepackage{upgreek}
\usepackage{comment}

\newcounter{rowcntr}[table]
\renewcommand{\therowcntr}{\roman{rowcntr}}
\newcolumntype{N}{>{(\refstepcounter{rowcntr}\therowcntr)}r}

\usepackage{multirow}
\DeclareSIUnit\bar{bar}
\DeclareSIUnit{\belmilliwatt}{Bm}
\DeclareSIUnit{\dBm}{\deci\belmilliwatt}

\arxivnumber{}

\title{RADES axion search results with a High-Temperature Superconducting cavity in an 11.7 T magnet}

\author [1]{S.~Ahyoune,}
\author [2]{A.~\'Alvarez Melc\'on,}
\author [*,1]{S.~Arguedas Cuendis,}
\author [3]{S.~Calatroni,}
\author [4]{C.~Cogollos,}
\author [2]{A.~D\'iaz-Morcillo,}
\author [4]{B.~D\"obrich,}
\author [5]{J.D. Gallego,}
\author [4]{J.M.~Garc\'ia-Barcel\'o,}
\author [6]{B.~Gimeno,}
\author [*,3,7]{J.~Golm,}
\author [8]{X. Granados,}
\author [8]{J. Gutierrez,}
\author [4,9]{L. Herwig,}
\author [10]{I.~G.~Irastorza,}
\author [8]{N. Lamas,}
\author [2]{A. Lozano-Guerrero,}
\author [11]{C.~Malbrunot,}
\author [3]{W. L. Millar,}
\author [1,12]{J.~Miralda-Escud\'e,}
\author [2]{P. Navarro,}
\author [2]{J. R. Navarro-Madrid,}
\author [8]{T. Puig,}
\author [13]{M. Siodlaczek,}
\author [8]{G. T. Telles,}
\author [3]{W.~Wuensch,}

\affiliation[1]{Institut de Ci\`encies del Cosmos, Universitat de Barcelona, 08028 Barcelona, Spain}
\affiliation[2]{Department of Information and Communications Technologies, Technical University of Cartagena, 30203 Cartagena, Spain}
\affiliation[3]{CERN - European Organization for Nuclear Research, Geneva, Switzerland}
\affiliation[4]{Max-Planck-Institut f\"ur Physik (Werner-Heisenberg-Institut), Boltzmannstr. 8, 85748
Garching bei M\"unchen, Germany}
\affiliation[5]{Yebes Observatory (IGN), 19141 Guadalajara, Spain}
\affiliation[6]{Instituto de Física Corpuscular (IFIC), CSIC-University of Valencia, 46980 Valencia, Spain}
\affiliation[7]{Institute  for  Optics  and  Quantum  Electronics,  Friedrich  Schiller  University  Jena, 07743 Jena,  Germany}
\affiliation[8]{Institut de Ciència de Materials de Barcelona (ICMAB), CSIC, 08193 Bellaterra, Spain}
\affiliation[9]{Technical University of Munich, Arcisstraße 21, 80333 Munich, Germany}
\affiliation[10]{Center for Astroparticles and High Energy Physics (CAPA), Universidad de Zaragoza, 50009 Zaragoza, Spain}
\affiliation[11]{TRIUMF, 4004 Wesbrook Mall, Vancouver, BC V6T 2A3, Canada and  Physics Department, McGill University, Montréal, Québec
H3A 2T8, Canada and Department of Physics and Astronomy, University of
British Columbia, Vancouver BC, V6T 1Z1 Canada }
\affiliation[12]{Instituci\'o Catalana de Recerca i Estudis Avan\c cats, 08010 Barcelona, Spain}
\affiliation[13]{Technical University of Darmstadt, Institute for Energy Systems and Technology, 64287 Darmstadt, Germany}

\affiliation[*]{Corresponding author}
\emailAdd{sergio.arguedas.cuendis@cern.ch; jessica.golm@cern.ch}

\abstract{We describe the results of a haloscope axion search performed with an \SI{11.7}{\tesla} dipole magnet at CERN. The search used a custom-made radio-frequency cavity coated with high-temperature superconducting tape. A set of \SI{27}{\hour} of data at a resonant frequency of around \SI{8.84}{\giga \hertz} was analysed. In the range of axion mass \SI{36.5676}{\micro \electronvolt} to \SI{36.5699}{\micro \electronvolt},  corresponding to a width of \SI{554}{\kilo \hertz}, no signal excess hinting at an axion-like particle was found. Correspondingly, in this mass range, a limit on the axion to photon coupling-strength was set in the range between g$_{a\gamma}\gtrsim$ \SI{6.3e-13}{\per \giga \electronvolt} and g$_{a\gamma}\gtrsim$ \textcolor{black}{\SI{1.59e-13}{\per \giga \electronvolt}} with a 95\% confidence level.}

\begin{document}
\preprint{CERN-EP-2024-076, MPP-2024-55}
\maketitle
\flushbottom

\section{Introduction}
\label{sec:introduction}
The QCD axion is a theoretically well motivated, hypothetical pseudoscalar particle beyond the Standard Model of particle physics. In addition to providing a solution to the strong CP problem \cite{Peccei:1977hh,Wilczek:1977pj,Weinberg:1977ma}, axions are an ideal candidate for dark matter \cite{Ipser:1983mw,Turner:1983sj, Preskill:1982cy, Abbott:1982af, Dine:1982ah}. If axions are created in a post-inflationary scenario, their present cosmic density and required mass, and their corresponding coupling to photons, can in principle be computed (see for example \cite{Klaer:2017ond}). Numerical simulations have indicated a mass of around a few tens of \SI{}{\micro \electronvolt}, with recent work pointing to higher masses \cite{Saikawa:2024bta}, although with substantial uncertainties related to the production of cosmic defects and possible non-standard expansion histories of the early Universe. This motivates the search presented here and numerous other set-ups. A full list of results of haloscope searches is maintained in \cite{ohare}.

Various experimental techniques have been developed in the search for QCD axions as dark matter candidates. Many rely on the inverse Primakoff effect \cite{Pirmakoff:1951pj}. Following the original haloscope proposal by P. Sikivie \cite{Sikivie:1983ip}, a radio-frequency (RF) cavity immersed in a strong external magnetic field  is used to detect a resonant photon produced through the interaction of the axion and magnetic fields. Led by the ADMX collaboration \cite{ADMX:2020ote}, the parameter space around a few \SI{}{\micro \electronvolt} has been probed down to the coupling values implied by the benchmark KSVZ and DFSZ models (see \cite{Irastorza:2018dyq} for a recent overview).

One of the main limitations of these techniques are the noise sources determining the effective temperature at which the resonant cavity signal can be extracted. This constrains the experimental sensitivity and the limiting value of the coupling constant $g_{a\gamma}$ that can be reached by the experiment. Many ideas have been developed to reduce the noise level and to increase the signal by maximising the magnetic field, cavity quality factor and other parameters \cite{ADMX:2021nhd,Adair:2022rtw,Backes:2020ajv, Alesini:2022lnp, Quiskamp:2022pks, Garcia-Barcelo:2023iri, Garcia-Barcelo:2023wrw}, leading to the exclusion of an impressive parameter space in the \SI{}{\micro \electronvolt} region.

To search at higher axion masses, resonant cavities become smaller as the inverse proportion of the mass, reducing the sensitivity to $g_{a\gamma}$ because of the smaller volume, and also because the  quality factor of the resonant cavity can be reduced with increasing frequency. Hence, other techniques have been proposed
above $\approx$ \SI{50}{\micro \electronvolt}, such as the MADMAX concept, a series of dielectric plates that can coherently radiate an axion field, assembled in a `booster' that amplifies the signal and enables an axion scan between \SI{50} and \SI{100}{\micro \electronvolt} \cite{Ivanov:2022hlb}. Another proposed experiment, ALPHA \cite{ALPHA:2022rxj}, aims to exploit plasma properties at relatively large axion masses.

The region between a couple of \SI{}{\micro \electronvolt} and \SI{50}{\micro \electronvolt} is a challenging one where cavities are reduced in size and therefore have lower quality factors due to worsened surface-to-volume ratio. In order to increase the volume at high resonant frequencies, the RADES collaboration (Relic Axion Dark matter Exploratory Set-up) has developed multi-cell cavities. A first axion search analysis was realised with a cavity array immersed in the \SI{9}{\tesla} CAST dipole magnet at CERN, and results were reported in \cite{CAST:2020rlf}.

Here we explore the prospects of increasing the quality factor of small cavities by using high-temperature superconducting (HTS) tapes, and present a search in the region centered around \SI{36.5687}{\micro \electronvolt}. The SM18 magnet at CERN, reaching a magnetic field value of \SI{11.7}{\tesla}, was used \cite{Magnet}. Although no frequency tuning was planned\footnote{Such tuning is utilised at CAPP-HeT-SC, see \cite{capptalk}.} in advance, a small amount of tuning occurred  owing to temporal pressure variations in the helium bath inside the magnet, which led to an effective tuning of our system over a \SI{312}{\kilo \hertz} frequency range.

A technical article on the performance of ReBCO (Rare-earth Barium Copper Oxide) tapes that were used has been previously published  \cite{Golm:2021ooj}. 
We note that this is not the first report of the use of superconductors to boost  quality factors of cavities: in \cite{Alesini:2019ajt} a NbTi cavity was used at \SI{2}{\tesla}, and the CAPP experiment also used YBCO tapes\footnote{Although, the CAPP cavities are designed for a solenoid magnet. They, therefore, use a different mode with a different geometric factor. The resonance frequency ($\sim$\SI{6.9}{\giga\hertz}) is lower than the RADES cavity ($\sim$\SI{8.8}{\giga\hertz}).} \cite{Ahn:2021fgb}. The novelty of our measurement lies in the fact that the data was taken in an \SI{11.7}{\tesla} magneto static field. 

In this work we also present an analysis pipeline that introduces new novelty elements like the usage of Principal Component Analysis to deal with the background noise produced by the electronics. This method showed promising results for non-standard background noise and has a lower impact on the axion line-shape compared to the standard Savitzky-Golay approach. Moreover, a modified Lorentzian analytical function was developed to better handle the power spectra produced with the new procedure.

This article is structured as follows: in section \ref{sec:setup}, we describe our measurement set-up and the concepts underlying the cavity design, and we provide the relevant cavity and environmental parameters.
Section \ref{sec:analy} details the data selection and the analysis procedure. Results of the axion search are given in section \ref{sec:res}. The conclusions are presented in section \ref{sec:conc}. 

\section{RADES HTS cavity and experimental set-up} 
\label{sec:setup}

\subsection{Cavity, HTS tapes and data acquisition}

As reviewed in \cite{AlvarezMelcon:2020vee}, previous results obtained in RADES relied on the use of cavities similar to microwave filters, i.e. with irises in between smaller sub-cavities. For this search, a quasi-cylindrical cavity shape was used. It was designed by adapting a rectangular cavity geometry by rounding the corners, allowing easy coating with the superconducting tapes. The left hand side (l.h.s.) of figure \ref{fig:JG_Cavity_design} shows a simulation of the surface currents flow in the cavity for the TE$_{111}$ mode (axion mode) of the cavity employed in this search. The  current flows are maximal along the long side of the cavity and minimal at the rounded ends of the cavity. The cavity is cut along the electrical field lines such that no currents flow across the gap of the cavity halves. The middle picture of figure \ref{fig:JG_Cavity_design} shows the cavity prototype assembled before coating, and the right image shows both cavity halves coated.
A first layer of copper coating was applied. A high temperature superconducting tape, ReBCO provided by the company THEVA \cite{PRUSSEIT2005866}, with the critical magnetic field above \SI{100}{\tesla}, was taped on the inner surface omitting the rounded ends. In this region, the surface currents are shallow, and having only copper in this region affects the cavity's quality factor by less than \SI{8}{\percent} {\cite{Golm:2021ooj}.

\begin{figure}[ht]
\centering
\includegraphics[width=1.0\textwidth]{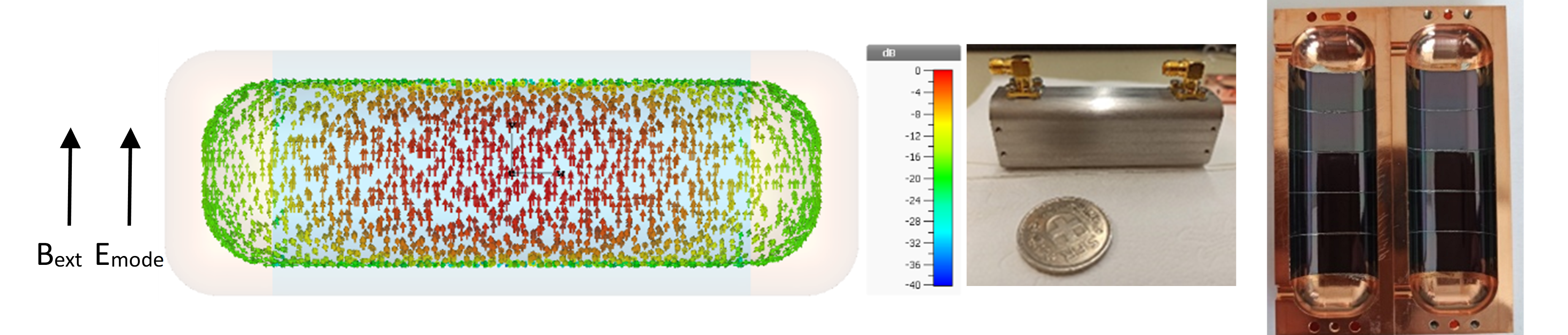}

\caption{\label{fig:JG_Cavity_design} 
Left: Direction of the surface currents (red showing the region of maximum surface current flow) for the TE$_{111}$ mode (axion mode) from CST Studio Suite® simulations. Center: The cavity assembly prototype uncoated and a coin for size comparison. Right: The copper coated cavity covered with ReBCO tape. The cavity body has an inner length of \SI{80}{\milli\meter}, a width of \SI{18.8}{\milli\meter}, and a height of \SI{24}{\milli\meter}. The corners are rounded with a radius of \SI{9}{\milli\meter}. }
\end{figure}

The coating of the cavity was performed at ICMAB \cite{ICMAB} in Barcelona where a method was developed to solder the tapes in the cavity with the substrate side facing outwards. Subsequent substrate removal exposed ReBCO to the RF field. The distribution of the superconducting ReBCO material in the cavity ensured that no RF currents were crossing the tapes. This method was previously tested on small flat samples and showed good results in performance compared to copper in magnetic fields up to \SI{9}{\tesla}, see \cite{Romanov:2020epk,Telles_2023}. 

The quality factor of the cavity coated with HTS tape improved by a factor of 1.5 compared to a copper cavity and was therefore used for data taking. This value was lower than predicted. Without magnetic field, we expected an improvement of Q$_0$ compared to the copper reference cavity by a factor of 5-6, but we only measured an improvement by a factor of 2 without magnetic field and 1.5 with magnetic field. The reason for the lower quality factor is the bending radius of the cavity geometry as it is smaller than the critical bending radius of the tape used. As we know, coated conductors lose their superconducting properties below the critical bending radius \cite{RomanovThesis, Otten_2016}. Studies to improve this value are ongoing. More details on the cavity design and coating method can be found in \cite{Golm:2021ooj}.

Here we summarise the parameters of the cavity design which are relevant to the axion search analysis, such as the  volume $V$ and the form factor $C$ representing the overlap of the magnetic field with the resonant mode \cite{CAST:2020rlf}. Both depend on the cavity geometry and are minimally affected by uncertainties like the thickness and quality of the soldering, fabrication errors, and coating thickness. The cavity consists of a 316LN stainless steel main body (with fabrication uncertainties \SI{20}{\micro \meter}) coated with a \SI{30}{\micro \meter} thick copper layer to which soldering tin and ReBCO tape (thickness \SI[separate-uncertainty = true]{40(5)}{\micro \meter}) were attached. Considering all these layers and the contraction of the stainless steel body by 0.3\% through a temperature change from \SI{300}{\kelvin} to \SI{4}{\kelvin} \cite{EKIN}, the volume was calculated to be V~=~0.0288~$\pm$~0.0002~L and the form factor C~=~0.634~$\pm$~0.001 using simulations in CST Studio Suite® \cite{CST}.

For data-taking, the cavity was connected to a cryogenic low noise amplifier from Low Noise Factory of type LNF-LNC6-20C \cite{lownoisefactory}, which was connected to the data acquisition system (DAQ) from TTI Norte \cite{TTI}. The DAQ consists of an analog and a digital stages. The analog part amplifies the input signal with a \SI{54.4}{\decibel} LNA and then converts it to an intermediate frequency centered at \SI{140}{\mega\hertz} using a Local Oscillator (LO). The digital stage has a bandwidth of \SI{12}{\mega\hertz} with a bin resolution of \SI{4577}{\hertz}, see \cite{CAST:2020rlf, SACThesis}.

\subsection{High field dipole magnet}

\begin{figure}[ht]
\includegraphics[width=1 \textwidth]{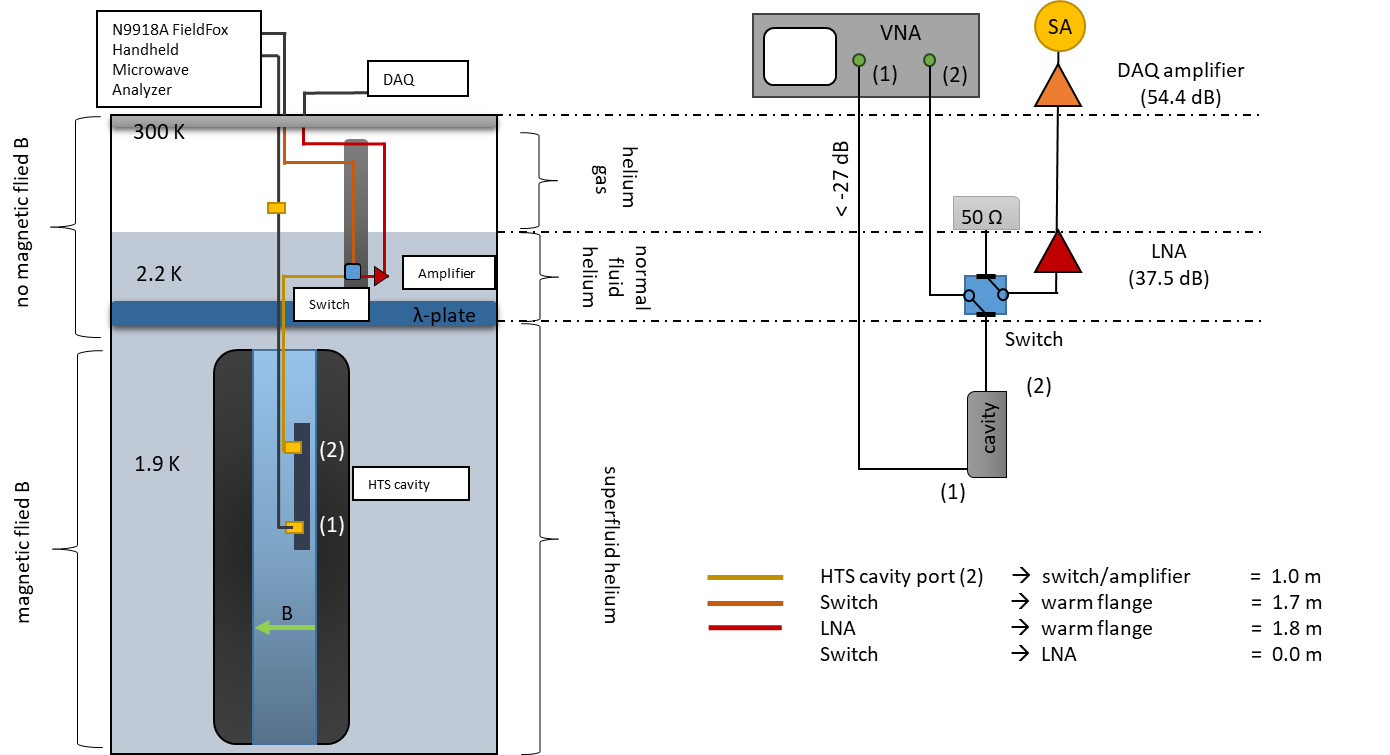}
    
    \caption{\label{fig:JG_SM18_set_up}} Left: Experimental set-up for RADES axion data-taking in a \SI{11.7}{\tesla} dipole magnet at CERN's SM18 hall in superfluid helium. Right: Wiring diagram of the set-up in the configuration for noise floor measurements to determine the noise temperature and for the quality factor measurements.  
\end{figure} 
The HTS cavity was immersed in the bore of a 2-m long dipole magnet in a single coil configuration which can reach magnetic fields higher than \SI{11}{\tesla}; for more details see \cite{Magnet}. Another copper-coated cavity of the same design was concurrently installed and served as a ``monitoring'' cavity.
A Hall sensor (Arepoc HHP- NPs) attached to the cavities assembly allowed the alignment of the cavities in the magnetic field before the cool-down of the magnet and measurements of the field during operation. One copper cavity surface was aligned with the Hall probe, and the HTS cavity was lined up with the copper cavity. At ambient conditions, the static magnetic field at a very low current was surveyed with the Hall probe to find the axis of the magnetic field. We expect less than 5 \% deviation of the field compared to the maximum B-field. This deviation is taken into account in the errors. 
During data taking, the magnet was kept at a current of I = \SI{11850}{\ampere}, corresponding to a magnetic field B = 11.7 $\pm$  \SI{0.1}{\tesla} which is constant along the cavity.
The magnet and the cavity assembly were inserted in a cryostat filled with liquid helium. The l.h.s. of figure \ref{fig:JG_SM18_set_up} shows the experimental set-up. The helium in the magnet bore was superfluid and the cavities were kept at a temperature of  T~=~1.897~$\pm$~\SI{0.008}{\kelvin} during data-taking. A Cernox resistance sensor was attached to each cavity for temperature measurement. The fluid around the cavity ensured the temperature was constant over the cavity length. 

During the data-taking period the pressure of the system fluctuated over a range of $p =$~1.27 to \SI{1.36}{\bar}, causing a slight variation of the helium relative dielectric constant, \\ $\varepsilon_r$~(\SI{1.9}{K},~\SI{1.36}{\bar})~=~1.058 (calculated from HEPAK\textsuperscript{TM} \cite{software_products}) of less than \SI{0.01}{\percent}. This resulted in a tuning of the resonant frequency over a range of \SI{312}{\kilo \hertz}. The pressure during the data taking depended on installations in the SM18 hall at CERN and the resulting environmental noise conditions, and was most stable during the weekend and at night, when most of the data were acquired. 

\subsection{Performance parameters}
\label{subsec:Performance parameters}
Besides volume and form factor, essential performance parameters for the axion search are the quality factor, coupling, amplification, and noise temperature. To enable a direct measurement of these parameters, a microwave switch was installed in the set-up.

The cavity has two ports, while one port is weakly coupled (terminated for the axion search - port (1)), we aim for critical coupling at the second port (strongly coupled - port(2)), i.e., a coupling between the cavity and the receiver chain of $\beta$ = 1.0. To minimise the attenuation of the signal coming from the cavity, the strongly coupled port was chosen to be as close as possible to the amplifier. For this reason, the top port in figure \ref{fig:JG_SM18_set_up} is the strongly coupled port. This port was connected to a switch by a copper RF cable of $l = 1.0 \pm 0.1$~m length. A cryogenic low-noise amplifier was connected directly to the switch allowing to bypass the amplifier to perform resonant frequency and quality factor measurements and obtain $Q_L$ and $\beta$. The quality factor and coupling were monitored on two subsequent days before the data-taking using a vector network analyser (VNA), this will be labelled as calibration data. By measuring the transmission coefficient $S_{21}$ a loaded quality factor of Q$_L$ = 36 656 $\pm$ \textcolor{black}{1832} was obtained\footnote{\textcolor{black}{A 5\% uncertainty was taken for $Q_L$ and $\beta$ to account for the systematic uncertainties of the system.}} and a coupling of $\beta$ = 0.81 $\pm$ \textcolor{black}{0.04} at \SI{11.7}{\tesla} by measuring the reflection coefficient $S_{11}$\cite{pozar2011microwave}. Both values remained constant within their error bars for a frequency range of \SI{295}{\kilo \hertz} during the period we monitored the Q-value. 

Figure \ref{fig:VNA_measurements} l.h.s shows one set of S-parameter of the calibration measurements that were performed in the experimental set-up in comparison to simulations. They are in good agreement, which was necessary since the form factor was calculated using simulations. Moreover, the neighbouring resonances are at \SI{7.97}{\giga\hertz} and \SI{9.05}{\giga\hertz}. Any type of mode mixing can be excluded. Figure \ref{fig:VNA_measurements}  r.h.s shows the unloaded quality factor versus magnetic field, which displays no hysteresis during ramp-up and ramp-down, demonstrating the robustness of the magnetic flux configuration at these high magnetic fields where the HTS is fully saturated.

\begin{figure}[htb]
      	
    \begin{minipage}[t]{0.5\linewidth}
	  \includegraphics[width=1\textwidth]{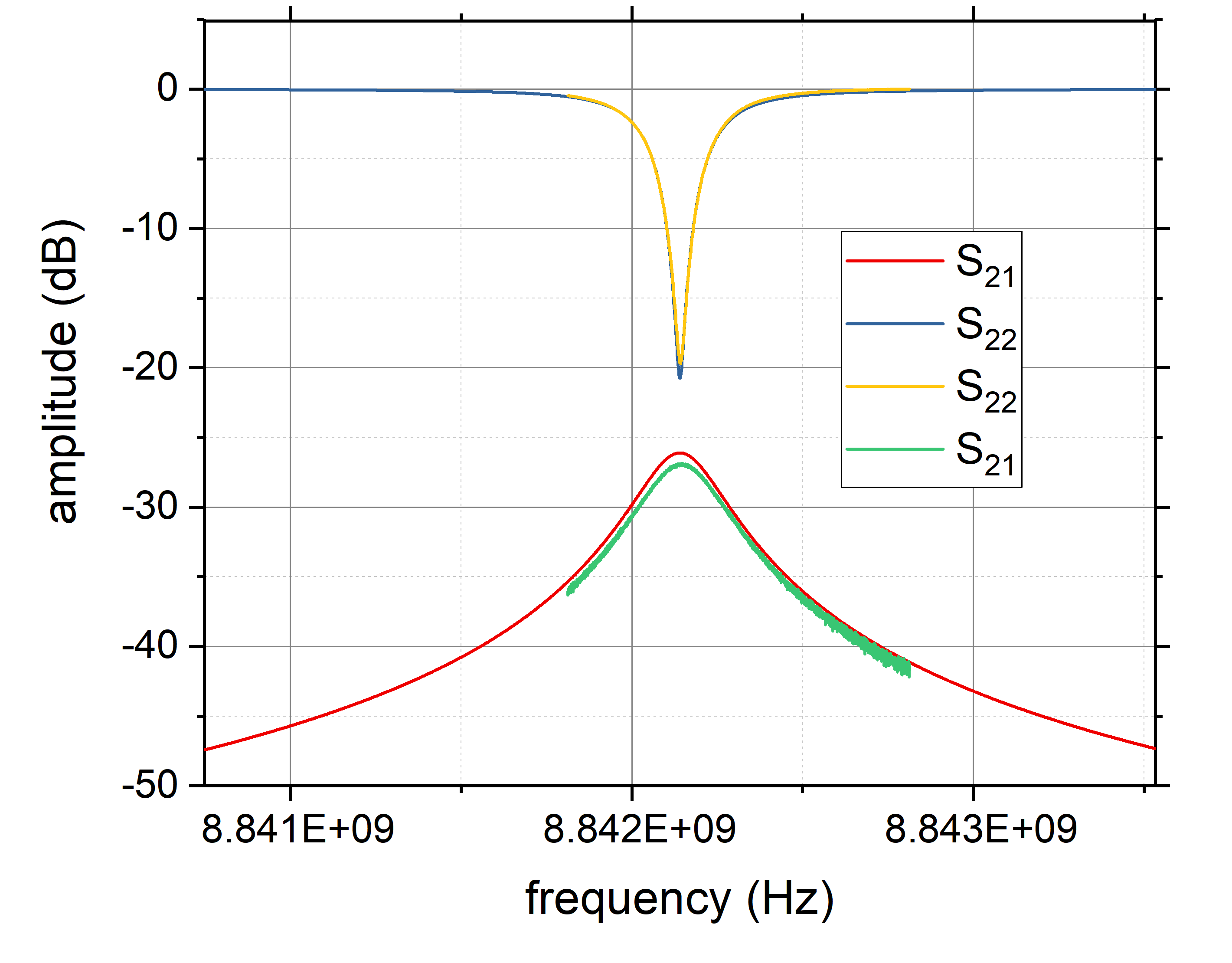}
    \end{minipage}%
  		\hfill
	\begin{minipage}[t]{0.5\linewidth}
	\includegraphics[width=1\textwidth]{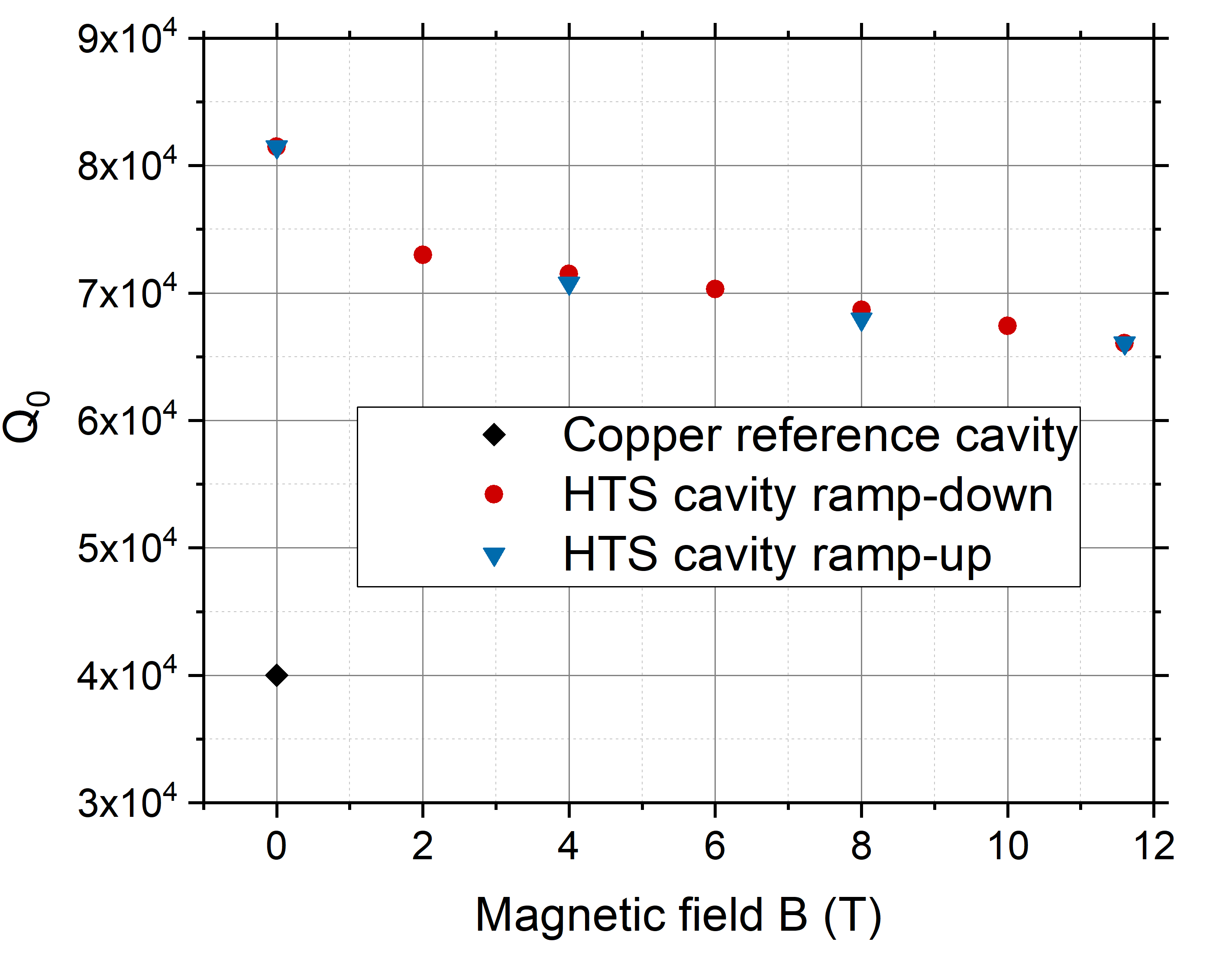}
	\end{minipage}
  		
 \caption{Left: Comparison of the measured S-parameters (transmission S$_{21}$ - green, reflection S$_{22}$ - yellow) and the simulated S-parameters (transmission S$_{21}$ - red, reflection S$_{22}$ - blue) of the HTS tape cavity. The measured S-parameters were used for quality factor and coupling calculation of the HTS cavity during the calibration measurements at \SI{1.9}{\kelvin} and \SI{11.7}{\tesla}. For the plot, cable losses were calibrated after the measurement. Right: Unloaded quality factor versus magnetic field for the HTS tape cavity at \SI{1.9}{\kelvin} for ramp-down and subsequent ramp-up in comparison to a copper coated reference cavity of the same shape at zero magnetic fields.}
 \label{fig:VNA_measurements}
\end{figure}

To determine the noise temperature of the system a direct power method was chosen. Therefore, it was important to know the gain of the low noise amplifier (LNA). Thanks to the switch, it was possible to inject an RF signal in the cavity and measure the outgoing power at the other port. This could be done by going through the amplifier once and by-passing it for a second measurement. Since the cables in both lines are almost of the same length (\SI{1.70}{\meter} for the bypassing line versus \SI{1.80}{\meter} for the amplifier line) it allowed good measurements of the gain, which was measured to be 37.4 $\pm$ \SI{0.9}{\decibel}. 

To measure the noise floor of the cryogenic low noise amplifier (LNA), the LNA was terminated with a \SI{50}{\ohm} cold load that was connected to the RF switch as shown in the wiring diagram in the right hand side (r.h.s.) of figure \ref{fig:JG_SM18_set_up}. For this measurement, another room temperature amplifier (\SI{54.4}{\decibel}) connected between the LNA and spectrum analyser (SA) had to be added to amplify the RF signal above the noise floor of the spectrum analyser. A wiring diagram for the noise floor measurement of the LNA, by-passing the cavity, is shown in the right hand side (r.h.s.) of figure \ref{fig:JG_SM18_set_up}. The noise floor was measured to be -71.1 $\pm$ \SI{0.2}{\dBm}. With this value the noise power $P_\text{amp}$ of the LNA taking into account the attenuation from the cables and the gain provided by the amplifiers was calculated \textcolor{black}{using equation \ref{eq:JG_Noisetemp2} to $P_\text{amp}$ = 157.8 $\pm$ \SI{1.9}{\dBm}} which corresponds to a noise power density of $P_\text{d}$ = \SI{0.05}{\dBm\per\hertz}.
\begin{equation}
\label{eq:JG_Noisetemp2}
    \textcolor{black}{P_\text{amp} [\SI{}{\dBm}]= \text{NF} -\text{Gain}_{\text{LNA}} - \text{Gain}_{\text{DAQ}} + \text{Cables}_{\text{cryo}} + \text{Cables}_{\text{warm}}}
\end{equation}
A list of the contributions for this calculation can be found in table \ref{tab:NT parameters}.}

\begin{table}[ht]
\begin{tabular}{ccc}
\toprule
Parameter & Value & Unit \\ \midrule
Noise floor  & -71.1 $\pm$  0.2 & \SI{}{\dBm}\\ 
Gain LNA  & 37.4 $\pm$  0.9 & \SI{}{\decibel} \\
Gain DAQ amplifier  & 54.4 $\pm$  0.3 & \SI{}{\decibel} \\\ 
Cryogenic cables  & -3.1 $\pm$  0.4 & \SI{}{\decibel}  \\\ 
Warm cables & -2.0 $\pm$ 0.1& \SI{}{\decibel} \\ \midrule

Noise power  $P_\text{amp}$   & \textcolor{black}{-157.8 $\pm$ 1.9} & \SI{}{\dBm}\\ 
\bottomrule
\end{tabular}
\centering
\caption{List of contributions used for noise power $P_\text{amp}$ calculation with direct noise measurement method}

\label{tab:NT parameters}
\end{table}

 The cable attenuation at \SI{300}{\kelvin} and \SI{10}{\giga\hertz} is quantified by the vendor as \SI{1.45}{\decibel \per \meter} \cite{datasheet_cables}. In this set-up, the cables have a temperature gradient from \SIrange{1.9}{300}{\kelvin} and the measurement was performed at frequencies below \SI{9}{\giga\hertz}. Therefore the cable attenuation is expected to be smaller than given in the datasheet, especially in the \SI{1.9}{\kelvin} region. During operation, the cable attenuation for the amplifier line could not be measured, but the S-parameters from the line cavity port (2) - switch - VNA gave an estimate of the attenuation. An attenuation of 3.1 $\pm$ \SI{0.4}{\decibel} for the \SI{3.7}{\meter} long cable chain was measured. The attenuation of the cable chain including the amplifier was estimated to have the same attenuation, which is a very conservative approach. 
Finally, the amplifier noise temperature was calculated using equations \ref{eq:JG_coversion} and \ref{eq:JG_Noisetemp}.

\begin{equation}
\label{eq:JG_coversion}
        \textcolor{black}{P_\text{amp} [\SI{}{\watt}] = 1/1000 \cdot 10^{\frac{P_\text{amp} [\SI{}{\dBm}]}{10}}}
\end{equation}

\begin{equation}
\label{eq:JG_Noisetemp}
       T_\text{amp}= \frac{P_\text{amp}}{k_{b}\cdot \Delta\nu}
\end{equation}

Where $k_{b}$ is the Boltzmann constant and $\Delta\nu$ the bandwidth corresponding to \SI{3}{\kilo\hertz} for our measurement. The noise temperature of the LNA was calculated to be $T_\text{amp}$~= 4.1~$\pm$ 0.8 K. This value is in good agreement with the noise temperature given in the datasheet of the amplifier (\SI{3.5}{\kelvin} at \SI{9}{\giga\hertz}) \cite{lownoisefactory}.

\textcolor{black}{To take into account the contribution of the cable from the cold attenuator to the amplifier (\SI{90}{\milli\meter} length, losses of \SI{0.2}{\decibel}), the formula for the noise temperature of the amplifier is given by:}

\begin{equation}
    \textcolor{black}{ T_\text{amp-new}= (L-1) T_\text{amb} + T_\text{amp} ,}
\end{equation}
\textcolor{black}{where $T_\text{amb}$ is the physical temperature of the cable (\SI{2.1}{\kelvin}), and $L$ is its linear loss factor (1.05). The new noise temperature of the amplifier would be \SI{4.2}{\kelvin}.}

To obtain the system noise temperature the contributions of the physical temperature of the cavity ($T_\text{cav}$ = \SI{1.9}{\kelvin}) and the thermal noise attenuated by \SI{27}{\decibel} injected through the weakly coupled port ($T_\text{coupling}$ = \SI{300}{\kelvin}/500 = \SI{0.6}{\kelvin}), need to be added to the amplifier noise temperature \cite{Y-method}. 
\begin{equation}
   \textcolor{black}{ T_\text{sys}= T_\text{amp-new} + T_\text{cav} + T_\text{coupling}}
\end{equation}
This results in a system noise temperature of \textcolor{black}{$T_\text{sys}$~=~\textcolor{black}{6.7} $\pm$ 0.8 K.~}\textcolor{black}{ The total noise temperature is primarily determined by the noise temperature of the cryogenic amplifier, as its high gain minimizes the contribution of subsequent stages.}

Due to the fact that access to the experimental facility was restricted in certain measurement intervals, it was necessary to measure the described performance parameters outside the data-taking interval.

\section{Selection and analysis of data}
\label{sec:analy}

\begin{figure}[ht]
\centering
\includegraphics[width=0.45 \textwidth]{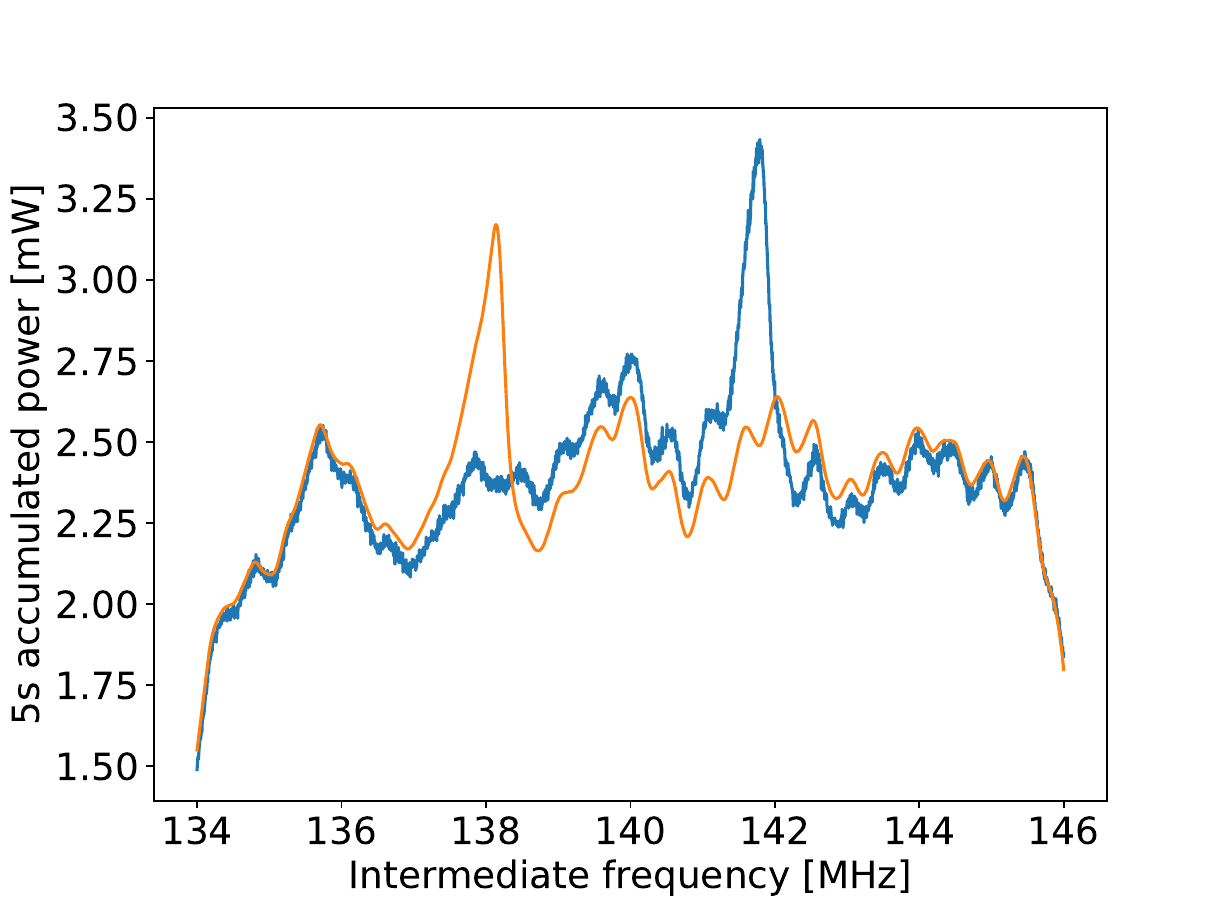}
\includegraphics[width=0.45 \textwidth]{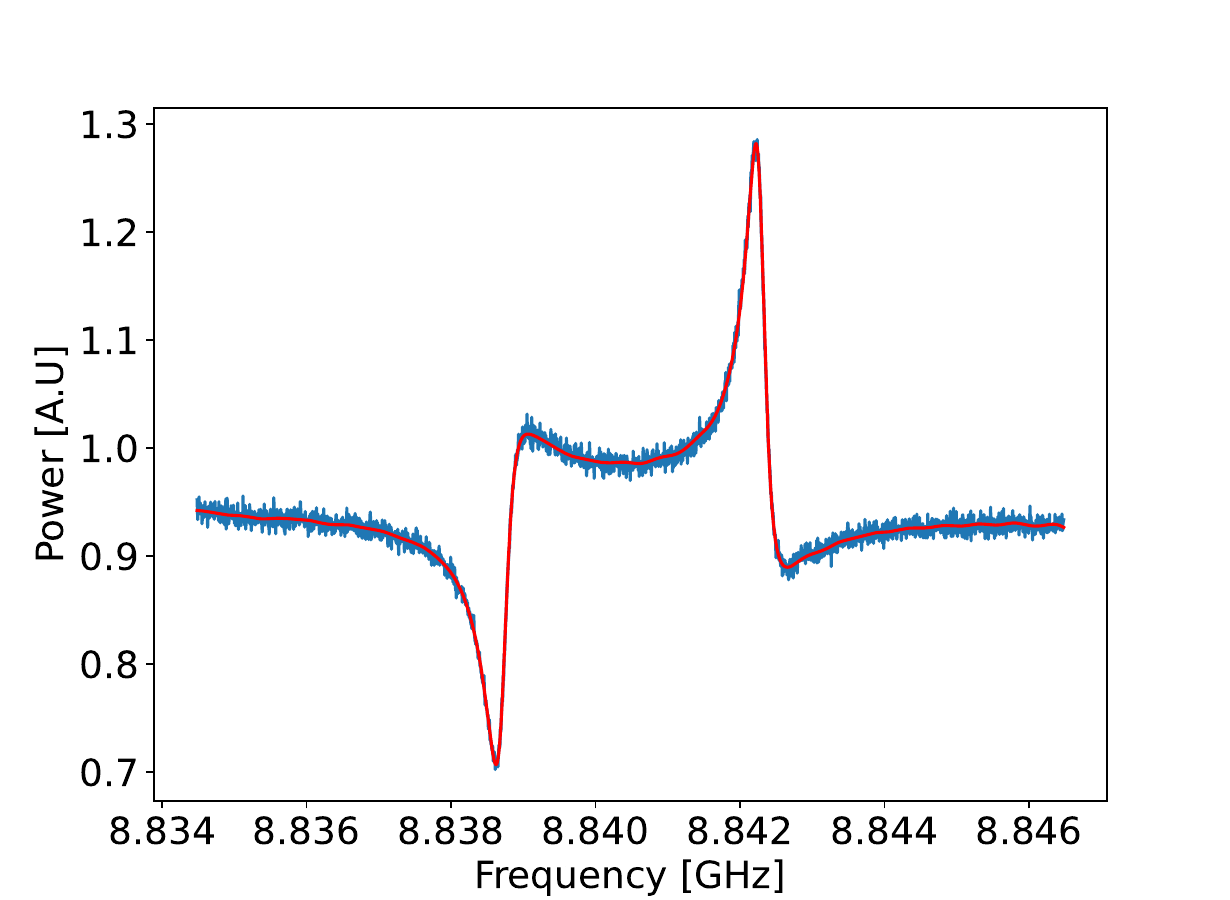}
	\caption{\label{fig:Spectra} 
Left: Typical spectra for two different LO frequencies ($l_1$  = \SI{8.700481644}{\giga \hertz} in blue and $l_2$ = \SI{8.704143756}{\giga \hertz} in orange). Upon changing the LO frequency, the cavity resonance peak changes position in the IF frequency. Right: In blue an example of a $\delta^d_{ik}$ spectrum obtained by the division of two spectra taken at two different LO frequencies. In red the fit function produced by equation \ref{eq:fit-function}.}
\end{figure}

The data used for this analysis were collected between the 13th and 15th of November 2021. A power spectrum was taken every t = \SI{5.36868}{\second} and consists of 2622 frequency bins in a \SI{12}{\mega \hertz} span centred at the intermediate frequency (IF) of \SI{140}{\mega \hertz}. The LO frequency can be set using the following equation:
\begin{equation}
\label{eq:LO-freq}
\text{IF} = x_0 - \text{LO},
\end{equation}
where $x_0 \approx \SI{8.842}{\giga\hertz}$ is the centre frequency of the cavity. The LO frequency was switched once per hour between two different LO frequencies ($l_1$  = \SI{8.700481644}{\giga \hertz} and $l_2$ = \SI{8.704143756}{\giga \hertz}). This was done in order to sample amplifier noise that was later used to remove part of the electronic background as described below. With these two LO values we made sure there was enough separation between the position of the cavity peak at the IF range as shown in the l.h.s. of figure \ref{fig:Spectra}. For this analysis we used approximately \SI{27}{\hour} at $l_1$ and \SI{22}{\hour} at $l_2$.

The analysis in this paper was designed following a special procedure with the goal of removing the electronic systematic of the read-out system. The analysis steps are outlined as follows:
\begin{enumerate}

    \item Division of the power spectra of $l_1$ by the average spectrum of $l_2$.
    \item With a DAQ calibration sample, characterise the cavity spectral profile, the electronic background (EB) and its time variability using Principal Component Analysis (PCA). 
    \item Remove the EB of the data-taking power spectra using a fit function based on a modified analytical Lorentzian function and the PCA. 
    \item Combine the individual spectra using a weighted sum to construct the Grand Unified Spectrum.
    \item Remove the remaining systematic structure using a Savitzky-Golay (SG) fit.
    \item Search for an axion by fitting its lineshape to the spectrum.
    \item If no signal is observed, a 95\% confidence level exclusion plot is created using Bayesian statistics.
    
\end{enumerate}  

Detailed description of steps 1, 5 and 6 can be found in \cite{CAST:2020rlf}. In this work we will focus on the improvements made over previous analysis to remove the EB and the systematic residual. 

In the off-the-shelf DAQs known to us, there is always a limit to the amount of hours that can be combined before a systematic residual becomes larger than the statistical noise fluctuations. Having a good control and understanding of this systematic residual is fundamental for the axion signal sensitivity to be limited by the statistical noise fluctuations rather than the EB.

Our \SI{12}{\mega \hertz} span for a single power spectrum is larger than the average span of other axion experiments which is between 1 and \SI{5}{\mega \hertz}. This allowed for the cavity response at the two different IF to be completely covered within these \SI{12}{\mega \hertz}, as shown in the l.h.s. of figure \ref{fig:Spectra}. This let us do the first step of the analysis pipeline. Moreover, $l_1$ and $l_2$ did not need to be tuned during the data-taking period because the tuning range of the cavity centre frequency was completely covered by the frequency span of the DAQ. 

However, it turned out that the custom-made electronics of our DAQ imprinted a characteristic wave-like EB (see the l.h.s. of figure \ref{fig:Spectra}) to the signal. The standard axion search analysis steps described in \cite{Brubaker:2017rna} could not account for that EB and the associated systematic residuals produced by the DAQ, in particular its time variations which prevent a simple calibration and removal. We therefore investigated other ways of handling the correction of these structures, and we opted for using PCA. 

To characterise this EB of the DAQ, we took calibrating data using a noise source as an input signal. The gain curve was computed using the following equation:

\begin{equation}
\label{eq:Gain-Curve}
    G = \frac{P_h - P_c}{T_h - T_c}~,
\end{equation}
where $P_h$ and $P_c$ are the powers measured by the DAQ when the noise source was on and off, respectively, and $T_h$ and $T_c$ are the noise equivalent temperatures for the noise source on and off, respectively.

\begin{figure}[ht]
\centering
\includegraphics[width=1 \textwidth]{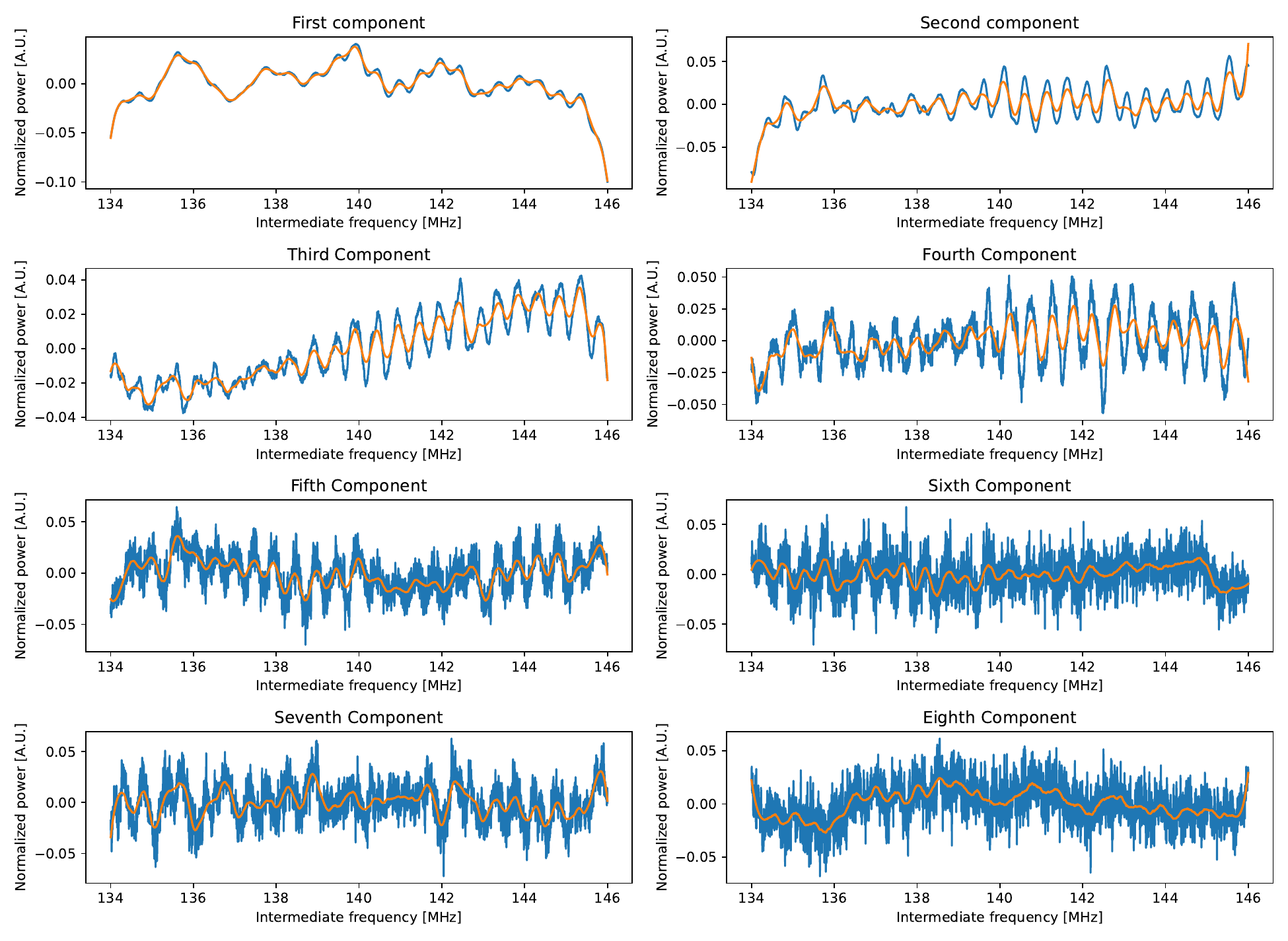}
\caption{\label{fig:PCA-components}}
In blue the first eight principal components of the DAQ gain curve. In orange the SG fit applied to them to smooth out the noise fluctuation. 
\end{figure}

A set of 13 gain curves was produced to perform the PCA, a statistical technique that performs a linear transformation of the original data into a new coordinate system where the variation of the data is described by its Principal Components \cite{PCA}. The gain curve measurements were done outside the SM18 environment after the data-taking period was finished. Figure \ref{fig:PCA-components} shows the PC of the EB. Time variations of the EB during the data-taking period can then be modelled as a linear combination of the PC. A SG filter \cite{savitzky64} (with W = 141 points and a polynomial degree N = 3) was applied to the PC to smooth out the noise fluctuations in the measurements to produce the gain curves. 
The optimum SG fit parameters (W, N) were identified by scanning the parameters and comparing the residuals distribution for each combination of parameters with an expected Gaussian distribution.

Note that the spectra used to create the PC and characterise the EB were produced using independent sets of data taken when the DAQ was isolated from the rest of the experimental set-up. This is an advantage compared to the previous axion search in RADES, where the EB is removed only with a SG fit applied directly on the data: the PCA helps remove systematics without causing any attenuation of any possible axion signal.

In ideal conditions, the resonance peak of a haloscope can be described using a Lorentzian function. However, our data could not be fitted in this way. Instead, we noticed the data could be fitted by a Lorentzian function multiplied by a linear term (see equation~\ref{eq:fit-function}). We attribute this modified cavity resonance profile to the presence of reflecting waves between the cavity and the low noise amplifier (LNA) due to the lack of a RF isolator between the two of them, although we were unable to positively identify the source of this modification. A linear term multiplying the Lorentzian function is used, which is found to fit the measured profile perfectly as far as the measurement noise allows us to tell. Following this approach, each spectrum obtained after performing step 1 of the analysis, denoted as $\delta^d_{ik}$, where \textit{i} represents the \textit{i}-th spectrum and \textit{k} the physical frequency (PH = IF + $l_1$), was fitted with the following function:
\begin{equation}
\begin{aligned}
f(x) &= \left(\left(1 + \frac{A_0(1-(\frac{x-x_0}{D_1}))}{1+(\frac{x-x_0}{D_1})^2}\right)/ \left(1 + \frac{A_1(1-(\frac{x-x_1}{D_2}))}{1+(\frac{x-x_1}{D_2})^2}\right)\right) \\
& \times (a_0 + a_1\cdot x + p_1\cdot g^\text{PC}_1(x) + p_2\cdot g^\text{PC}_2(x) + ... + p_7\cdot g^\text{PC}_7(x))~,\\
\end{aligned}
\label{eq:fit-function}
\end{equation}
where the first two terms describe the modified Lorentzian functions measured at the two LO frequencies. The amplitudes are given by $A_0$ and $A_1$, $x_0$ and $x_1$ are the centre frequencies for the two LOs, and $D_1$ and $D_2$ are their full widths at half maximum. The $g^\text{PC}_s(x)$ (s = 1,...,7) are the SG fits of the PC (see orange line of figure \ref{fig:PCA-components}). The $p_s$ represent the weights given to each PC.

To make sure that the fit gave meaningful values for $x_0$ and Q-loaded, the calibration measurements (see figure \ref{fig:VNA_measurements}) were used to compare the values measured by the VNA (which are described by a standard Lorentzian function with a well known physical model) with the ones given by the fit function. The centre frequency $x_0$ was similar for both Lorentzian models, while the Q-value of the modified Lorentzian was higher by an offset of around 9500. The Q-values given by the fit during the calibration were similar within uncertainties to the ones measured during the data-taking period. Since the offset is known, the real physical Q-loaded during data-taking corresponds to the value measured with the VNA during the calibration and this value was the one used for the exclusion limit described below.

The r.h.s. of figure \ref{fig:Spectra} shows a fit function on top of one $\delta^d_{ik}$ spectrum.  The $\delta^d_{ik}$ spectra are then divided by their corresponding fit function and centred at zero:
\begin{equation}
\label{eq:5s-norm-spec}
    \delta^n_{ik} = \frac{\delta^d_{ik}}{f_{ik}}-1.
\end{equation}

\begin{figure}[ht]
\centering
\includegraphics[width=0.45 \textwidth]{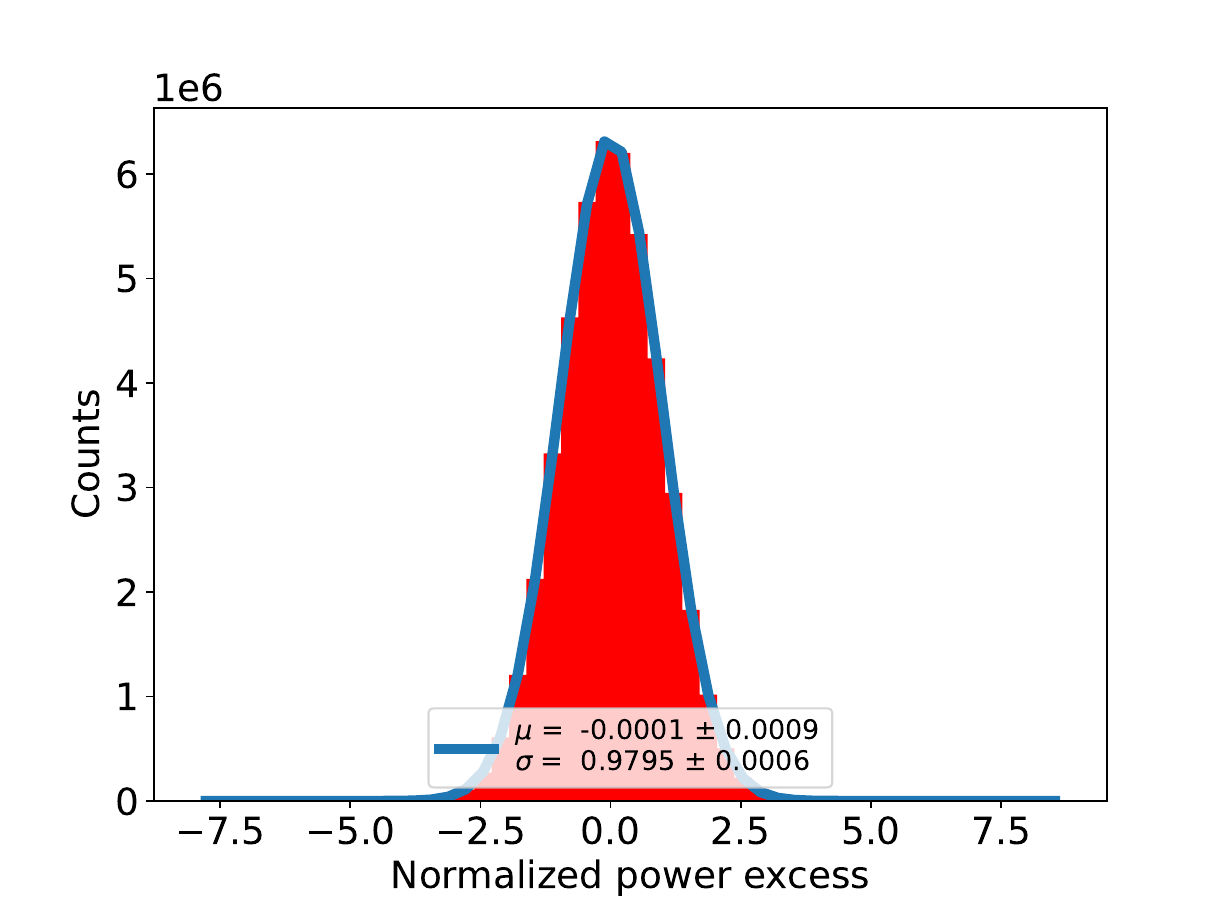}
\includegraphics[width=0.45 \textwidth]{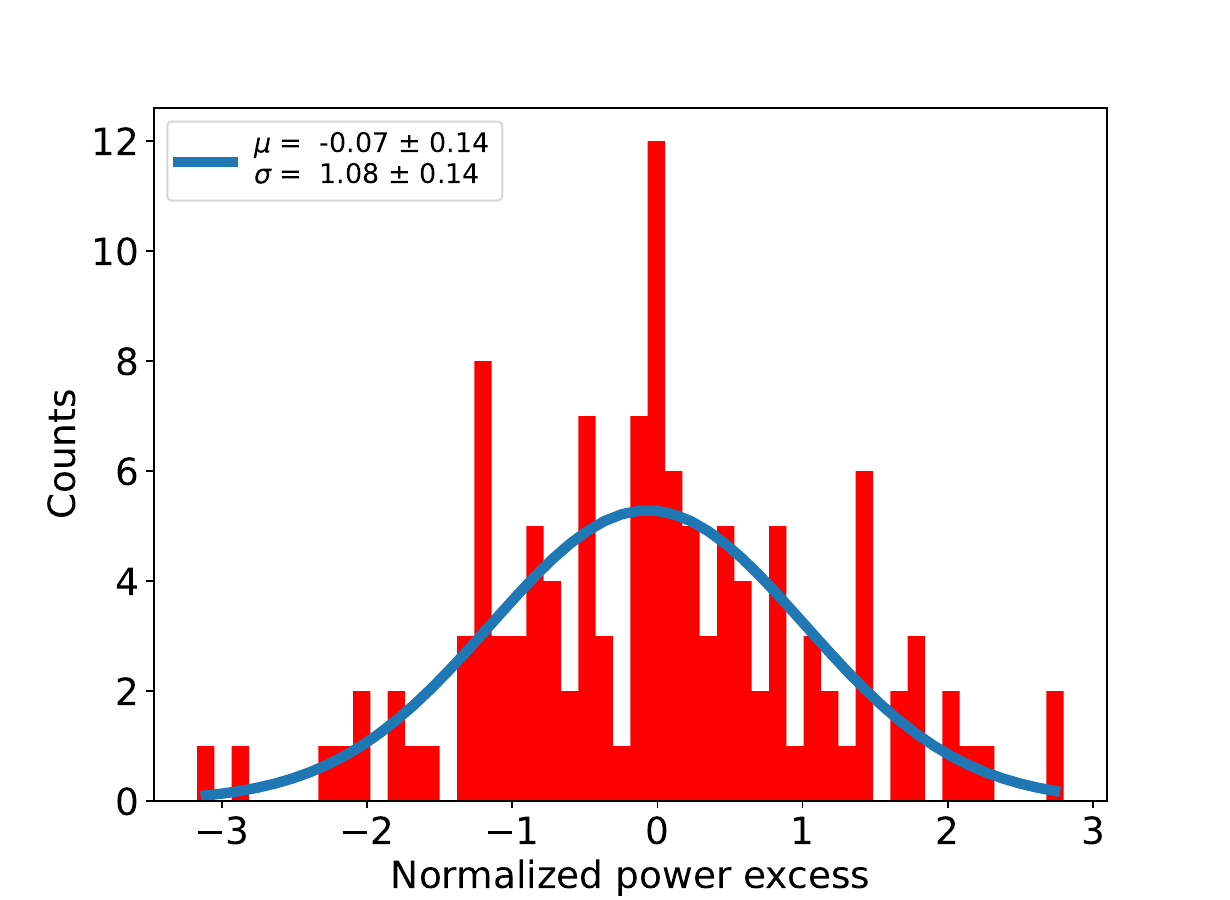}
\caption{\label{fig:Histo}}
Left: Histogram for all $\delta_{ij}^{n}$/$\sigma$. Right: Normalised histogram of the weighted GUS.
\end{figure}

If the fit is able to remove the EB, the $\delta^n_{ik}$ spectra should be samples of a Gaussian distribution centred at $\mu = 0$ and standard deviation $\sigma^{t} = 1/\sqrt{\Delta\nu \cdot t}$, where $\Delta \nu =$\SI{4578}{\hertz} is the resolution bandwidth of a single bin and $t =$ \SI{5.36868}{\second} is the integration time of each spectrum. The l.h.s. of figure \ref{fig:Histo} shows the histogram of all $\delta^n_{ik}$/$\sigma^{t}$. 


As described in section \ref{sec:setup}, pressure changes inside the cryostat resulted in a tuning range of \SI{312}{\kilo \hertz}\footnote{Almost 85 \% of the time was taken within a narrower effective tuning range of \SI{55}{\kilo \hertz}.}. To obtain an adequately weighted spectrum accounting for this tuning, the spectra were re-scaled by Lorentzian functions denoted $L_{ik}$\footnote{Note that the Lorentzian $L_{ik}$ is normalised to 1 at $x=x_0$, to match the axion power $P_a$ given in equation \ref{eq:axion power} at the resonance frequency.}
\begin{equation}
\label{eq:Lorentz-fit-function}
    L_{ik}(x)= 
\frac{1}{1 + 4Q_L^2(x/x_0-1)^2} ~,
\end{equation}
where the centre frequency used is the $x_0$ value given by the fit to each individual spectrum (see equation \ref{eq:fit-function}), and $Q_L$ is the loaded quality factor, with the value of $Q_L=36656$ measured with the VNA during the calibration measurements done before the data-taking period as detailed in section \ref{sec:setup}. The re-scaled $\delta^s_{ik}$ and standard deviation $\sigma^s_{ik}$ were computed as:

\begin{equation}
\label{eq:delta-s}
\delta^s_{ik} = \frac{\delta^n_{ik}}{L_{ik}}~,
\end{equation}

\begin{equation}
\label{eq:sigma-s}
\sigma^s_{ik} = \frac{\sigma^n_{i}}{L_{ik}}~.
\end{equation}
Here, the noise of each spectrum $i$, $\sigma_i^n$, is not the theoretical value $\sigma^{t}$, but is instead computed directly from the dispersion of the frequency bins in each spectrum; in practice there are small jumps in this dispersion at the level of $\sim$ 2\% caused by small systematics, but in any case changing this noise does not change our results significantly.

\begin{figure}[ht]
\centering
\includegraphics[width=0.45 \textwidth]{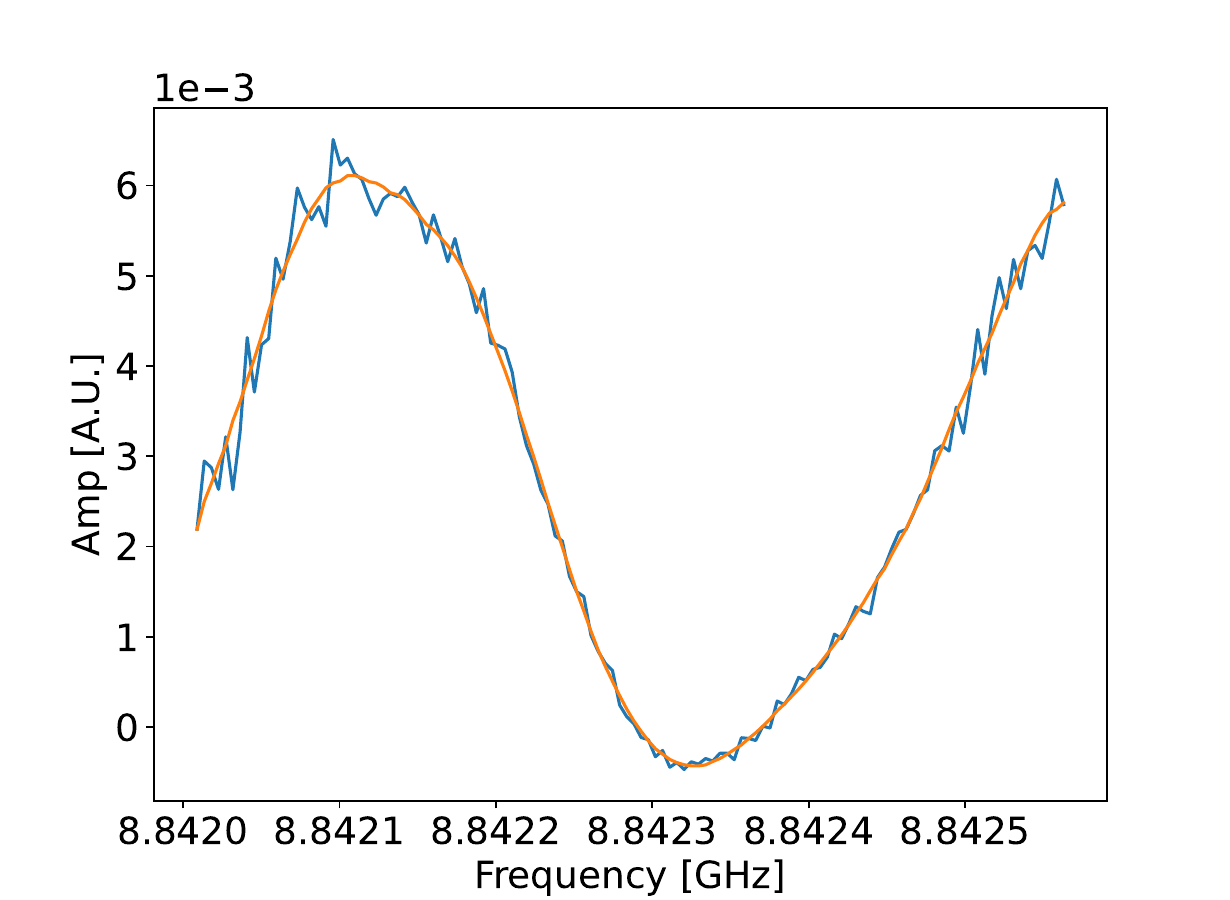}
\includegraphics[width=0.45 \textwidth]{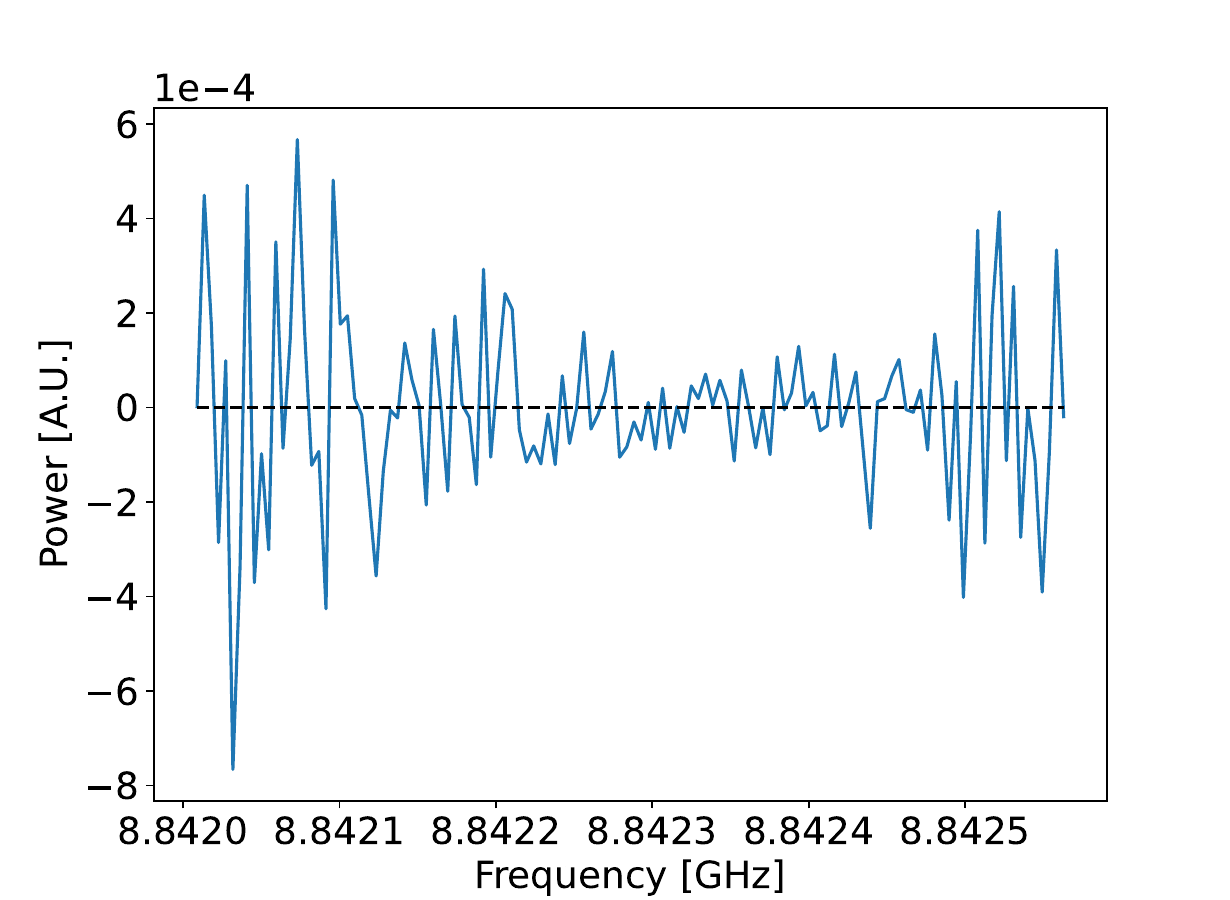}
\caption{\label{fig:Systematic-residual}}
Left: In blue the weighted GUS $\delta^g_{k}$. The distribution is inconsistent with a statistical noise distribution. The SG fit appears in orange. Right: Final weighted GUS $\delta^f_{k}$ resulting from the subtraction of the left spectrum by its SG fit.
\end{figure}

The weighted Grand Unified Spectrum (GUS), shown in the l.h.s. of figure \ref{fig:Systematic-residual} can be computed using the weighted sum of the power spectra \cite{Brubaker:2017rna}:
\begin{equation}
\label{eq:GUS}
\delta^g_{k} = \sum_i \delta^s_{ik} \cdot w_{ik},
\end{equation}
where the weights $w_{ik}$ are given by
\begin{equation}
\label{eq:weights}
w_{ik} = \frac{(\sigma^s_{ik})^{-2}}{\sum_{i^\prime} (\sigma^s_{i^\prime k})^{-2}}.
\end{equation}

The standard deviation of the weighted GUS spectrum is similarly computed as
\begin{equation}
    \sigma_k^g= \sqrt{\sum_{i} (\sigma^s_{i k})^{2}\cdot w_{ik}^2}.
\end{equation}

However, $\sigma_k^g$ needs to be corrected due to the fact that on the first analysis step the spectra are divided by the average spectrum of $l_2$, which has its own noise fluctuation $\sigma^{l_2} = 1/\sqrt{\Delta\nu \cdot t^{l_2}}$, where $t^{l_2}=$  \SI{22}{\hour} is the total time taken for the spectra with local oscillator $l_2$. The corrected standard deviation is given by:

\begin{equation}
    \sigma_k^{g*} = \sqrt{(\sigma_k^g)^2 + (\sigma^{l_2})^2}.
\end{equation}

For the frequency range of interest (see section \ref{sec:res}), we find that the GUS that is obtained with this procedure still contains a systematic which dominates over the statistical fluctuations, originating from residual EB (see l.h.s. of figure \ref{fig:Systematic-residual}). A SG fit (labelled $\delta^{\text{SG}}_k$) with W = 27 and N = 3 was done to characterise this residual. The residual was removed to create the final weighted GUS ($\delta^f_k = \delta^g_k - \delta^{\text{SG}}_k$). The r.h.s. of figure \ref{fig:Systematic-residual} shows the result after the systematic residual was removed. The histogram of the normalised spectrum ($\delta^f_k/\sigma^{g*}_k$) exhibits the expected Gaussian distribution (see r.h.s. of figure \ref{fig:Histo}).

This procedure induces an attenuation and  distortion of the axion signal (see section \ref{sec:res}), but the effect is smaller than the one evaluated in our previous analysis \cite{CAST:2020rlf}, where the SG fit was done with W = 15 and N = 3. A larger window moves the cut off frequency of the SG filter to lower frequencies reducing the attenuation of the axion signal \cite{SACThesis}, an improvement that is obtained thanks to the PCA that we have applied. 
Moreover, with this new analysis approach the EB and systematic residual removal was achieved for the whole frequency span of our DAQ. This allows the analysis procedure to work on a broader frequency range. 

\section{Results}
\label{sec:res}

We now proceed to analyse the section of the GUS that has the highest signal-to-noise ratio and is therefore most interesting to limit any possible presence of an axion signal. We choose a frequency interval to cover the tuning range of the central frequency of the cavity in the data set that we have analysed, plus the frequency interval for which the sensitivity drops by \SI{3}{\decibel} from the maximum of the Lorentzian peak, on either side. This corresponds to a total frequency interval of \SI{554}{\kilo \hertz}, from approximately \SI{8.84201}{\giga\hertz} to \SI{8.84256}{\giga\hertz}, shown in the r.h.s. of figure \ref{fig:Systematic-residual}.

\begin{figure}[ht]
\centering
\includegraphics[width=0.45 \textwidth]{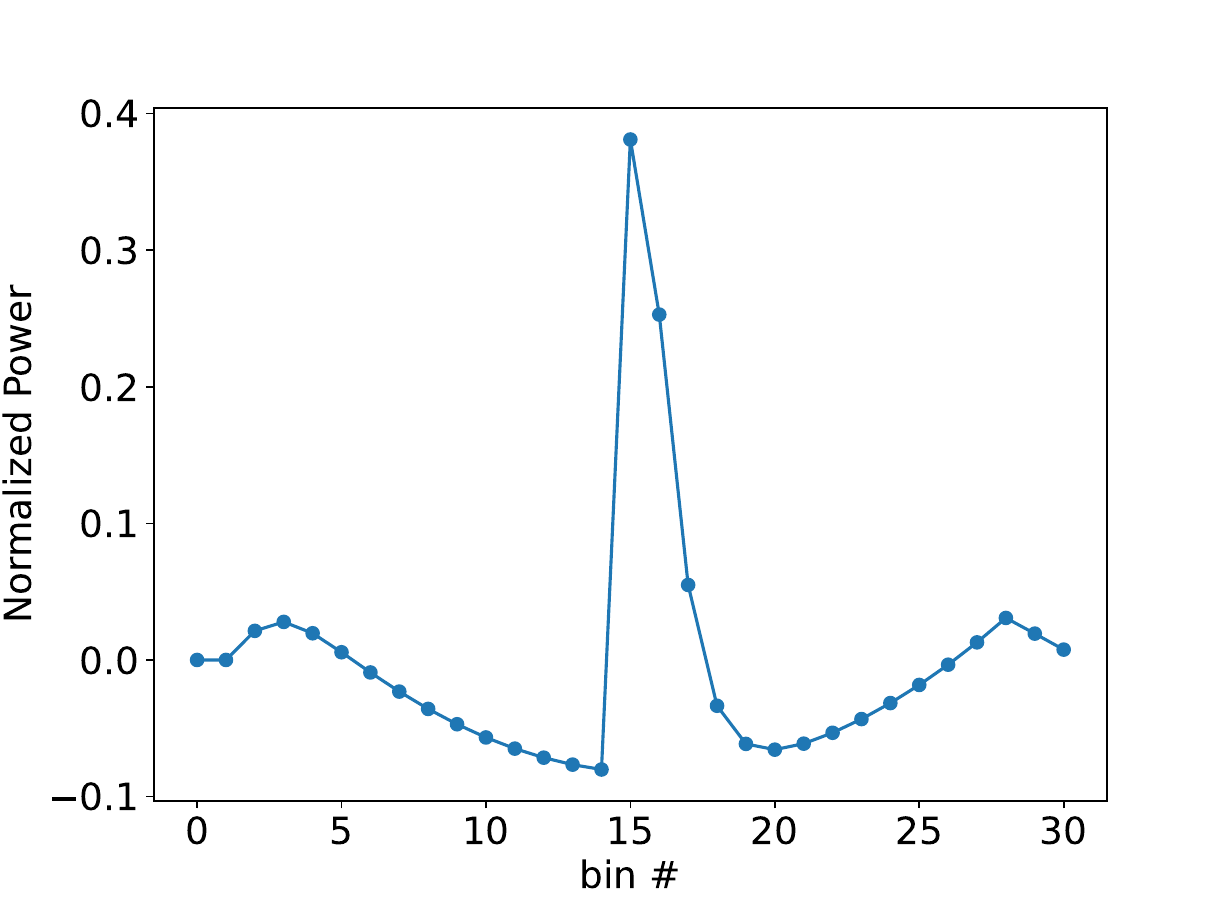}
\includegraphics[width=0.45 \textwidth]{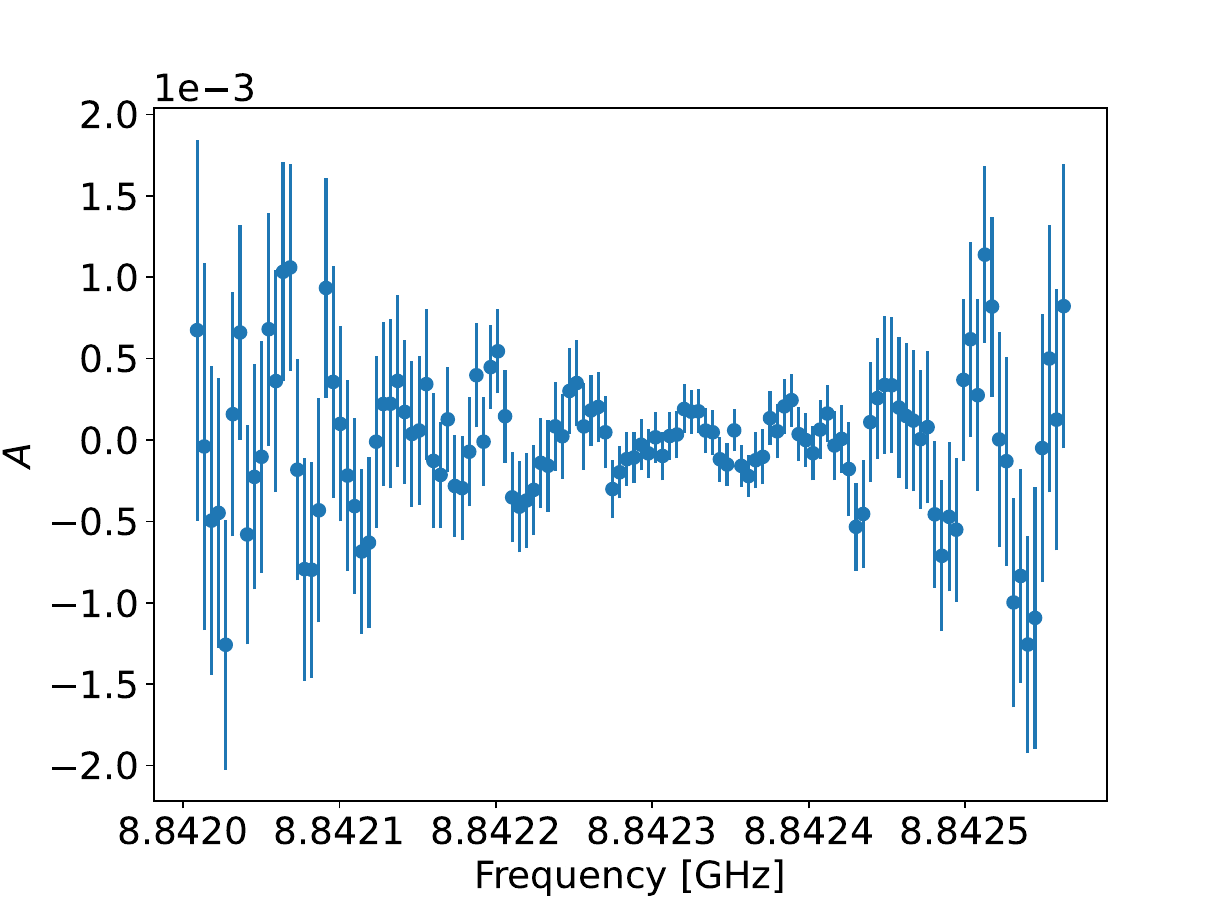}
\caption{\label{fig:Amp_plot}}
{\it Left:} Normalised power of the distorted and attenuated axion line shape. {\it Right:} Amplitude $A$ (in units of normalised power excess) of the axion signal provided by the
fit for each of the probed axion frequencies.
\end{figure}

To test for the presence of any possible axion signal with the theoretically expected form of the axion line shape for our Galactic halo, we use a discretised version of the modelled axion line shape \cite{Brubaker:2017rna}:
\begin{equation}
    \label{eq:Line-Shape-p}
    D_q = \int_{\nu_a + (q-1)\Delta\nu}^{\nu_a + q\Delta\nu} f(\nu)d\nu,
\end{equation}
where $\nu_a$ is the axion frequency, $q$ labels the bin number, $\Delta\nu$ the resolution bandwidth and $f(\nu)$ is the normalised frequency line profile based on the standard isothermal spherical dark matter halo model \cite{Turner:1990qx, Brubaker:2017rna}. We then apply the same SG filter that is applied to the data to remove the systematic electronic residual in order to obtain the attenuated and distorted axion line shape (as described in \cite{SACThesis}), $D_q^d$. The resulting model line profile is shown in the l.h.s. of figure \ref{fig:Amp_plot}. 

This model fit function,
\begin{equation}
    \label{eq:fit-function-p}
    y = A \cdot D^d_q ~,
\end{equation}
where $A$ (the only free parameter in the fit) is the amplitude of the axion signal, is then used to search for an axion signal on the weighted GUS.

The r.h.s. of figure \ref{fig:Amp_plot} shows the value of $A$ obtained at each frequency step of width $\Delta\nu$. The largest excess in the amplitude plot has a 2.10$\sigma$ local significance. 
A search including a misalignment ($\delta\nu_a$) within a bin width between the axion frequency $\nu_a$ and the DAQ $k$ frequencies yielded the highest outlier at a local significance of 2.12$\sigma$ only. Thus no significant signal above statistical fluctuations was observed.

The amplitude of the upper limit ($A_{\text{UL}}$) was computed using Bayesian statistics \cite{CAST:2020rlf}:
\begin{equation}
    \label{eq:Upper-Limit-p}
    \frac{1}{N}  \int_0^{A_{\text{UL}}} \text{e}^{-(\sum_k S(\delta_{k}^{f},A)/2)} dA = 1 - \alpha,
\end{equation}
 where $N$ is the normalization factor, $S(\delta_{k}^{f},A)$ is the $\chi^2$ function (both given in \cite{CAST:2020rlf}) and 1-$\alpha$ is the credibility level (CL) of the upper limit. For this analysis a 95$\%$ CL was used.

 By multiplying the $A_{\text{UL}}$ values by the noise power ($P_N = k_b T_{\text{sys}} \Delta\nu$), we convert them into power values. An exclusion limit is established using:
\begin{equation}
\label{eq:SNR}
    A_{\text{UL}} \cdot {P_N} = P_a 
\end{equation}
and the conversion power $P_a$ is given by:
\begin{equation}
\label{eq:axion power}
    P_a = g_{a\gamma}^2\rho_a \frac{1}{m_a}B^2 V C \frac{\beta}{1+\beta}\text{min}(Q_L, Q_a)\eta,
\end{equation}
where $g_{a\gamma}$ is the axion-photon coupling, $\rho_a= \SI{0.45}{\giga \electronvolt \per \cubic \centi \meter}$ \cite{Read:2014qva} is the local dark matter density, $m_a$ is the axion mass, $B$ is the magnetic field strength, $V$ is the volume of the cavity, $Q_L$ and $Q_a$ are the loaded cavity and axion quality factor respectively, $C$ is the geometric factor, $\beta$ is the coupling coefficient between the cavity and the receiver chain and $\eta$ is the attenuation factor between the cavity output port and the LNA, which was estimated from the cable data sheet. Table \ref{tab:RADES parameters} summarises the values used for this analysis. Note the largest error in these experimental parameters affecting our upper limits is that in $T_{\rm sys}$, and that we do not include any error in the DM density $\rho_a$.

\begin{figure}[ht]
\centering
\includegraphics[width=1 \textwidth]{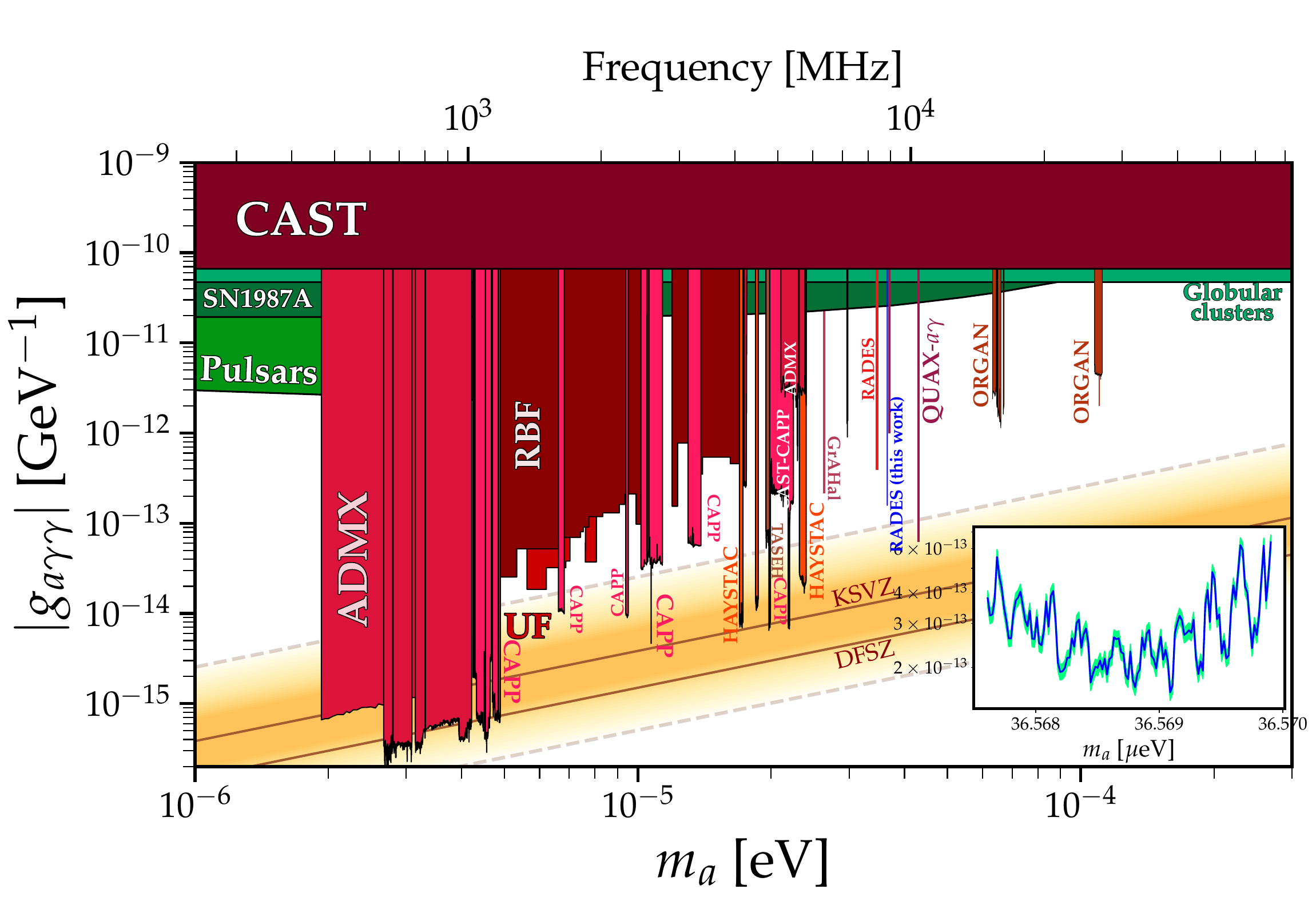}
\caption{\label{fig:Exclusion-Limit}
Axion mass versus axion-photon coupling phase-space. In blue the RADES axion-photon coupling exclusion limit with 95$\%$ credibility level presented in this article. Other haloscope results: RBF \cite{DePanfilis:1987dk}, UF \cite{Hagmann:1990tj}, ADMX and ADMX-SideCar \cite{ADMX:2021nhd,Braine:2019fqb,Du:2018uak,Boutan:2018uoc}, CAPP \cite{Adair:2022rtw, Lee:2020cfj,Jeong:2020cwz}, HAYSTAC \cite{Zhong:2018rsr,Backes:2020ajv}, QUAX \cite{Alesini:2019ajt,Alesini:2020vny, Alesini:2022lnp} and ORGAN \cite{McAllister:2017lkb,Quiskamp:2022pks} and the CAST solar axion results \cite{Anastassopoulos:2017ftl} are plotted for comparison, see \cite{ohare} for a full list of references and the raw source for the plot. Inset: Zoom-in of the parameter range probed in this work (\SI{36.5676}{\micro \electronvolt} $< m_a <$ \SI{36.5699}{\micro \electronvolt}), where the green region represent the uncertainty of the measurement. The l.h.s. of the spectrum shows less sensitivity because fewer data were recorded at those frequencies.}
\end{figure}

\renewcommand{\arraystretch}{1.3}
\begin{table}[ht]
\begin{tabular}{ccc}
\toprule
Parameter & Symbol & Value\\ \midrule
Axion DM Density & $\rho_a$ & \SI{0.45}{\giga \eV \per \cubic \centi \meter} \\ 
Total cavity volume & $V$        & \SI[separate-uncertainty = true]{0.0288(2)}{\liter}\\
Magnetic field & $B$        & \SI[separate-uncertainty = true]{11.7 (0.1)}{\tesla}\\ 
Loaded quality factor & $Q_L$  & \textcolor{black}{\SI[separate-uncertainty = true]{36656(1832)}{}}\\ 
Coupling factor & $\beta$  & \textcolor{black}{\SI[separate-uncertainty = true]{0.81(0.04)}{}}\\ 
Form factor & C        & \SI[separate-uncertainty = true]{0.634(1)}{}\\ 
Cable attenuation & $\eta$    & \SI[separate-uncertainty = true]{0.85(4)}{}\\ 
Noise Temperature & $T_\text{sys}$      & \textcolor{black}{\SI[separate-uncertainty = true]{6.7(0.8)}{\kelvin}}\\ 
\bottomrule
\end{tabular}
\centering
\caption{List of the environmental parameters used for the computation of the upper limit on the axion-photon coupling.}
\label{tab:RADES parameters}
\end{table}
An exclusion limit between g$_{a\gamma}\gtrsim$ \SI{6.3e-13}{\per \giga \electronvolt} and g$_{a\gamma}\gtrsim$ \\textcolor{black}{\SI{1.59e-13}{\per \giga \electronvolt}} was obtained for the range \SI{36.5676}{\micro \electronvolt} $< m_a <$ \SI{36.5699}{\micro \electronvolt}.
The error propagation of the experimentally determined physical quantities included in equation \ref{eq:axion power} is used to determine the magnitude of the systematic uncertainties. These uncertainties are represented as the green band in the insert of figure \ref{fig:Exclusion-Limit}. They were disregarded in the estimation of the exclusion limit since they account for less than 10\% of the error. 

Figure \ref{fig:Exclusion-Limit} shows the result of this analysis in the context of other haloscope searches: A competitive limit is set by RADES at an axion mass slightly lower than the QUAX results \cite{Alesini:2019ajt,Alesini:2020vny, Alesini:2022lnp}  and above the highest {ADMX-Sidecar} limit \cite{Boutan:2018uoc}.

\section{Conclusions} 
\label{sec:conc}
This work presents the analysis and results of the first data-taking campaign of the RADES group with a superconducting RF cavity.
The use of a ReBCO taped cavity increased the quality factor compared to an usual copper coating by \SI{50}{\percent}. Other studies have shown that this type of superconducting coating can yield an even higher increase of the quality factor, but this was limited here due to the 9 mm curvature radius of the cavity corners. \textcolor{black}{There is a pathway we have started to overcome these limitations. Firstly, we are changing the tape provider to Fujikura, which has a critical bending radius smaller than the bending radius of the cavity. Together with an} upgraded coating procedure for a new cavity we demonstrated a quality factor improvement by a factor of five with no magnetic field applied
\cite{PBCtalk_JGolm}. \textcolor{black}{Secondly, if one wants to reach even higher quality factors, there are two main aspects that have to be replicated. (a) To operate in a Lorentz force-free configuration, that is where the B-field is parallel to most of the tape plane. (b) Avoid curved cavities as much as possible. In this regard, a polyhedral cavity that fulfils these two conditions has been designed for measurements in solenoid magnets. Thirdly, other steps will need to be done to reach the highest possible quality factors, such as polishing the sides of each facet of the polyhedral cavity to properly align them. }

No signal excess has been found in the data analysed here when fitting the line-shape of a conventional axion dark matter halo model, in the range (\SI{36.5676}{\micro \electronvolt} $< m_a <$ \SI{36.5699}{\micro \electronvolt}) using \SI{27}{\hour} of data. The obtained upper limit for the axion-photon coupling in our narrow  frequency range is below the prior CAST limit \cite{Anastassopoulos:2017ftl} by more than two orders of magnitude. In addition, it improves the previous RADES search, which was made at a slightly lower cavity frequency, by a factor of 2.5 in the sensitivity to $g_{a\gamma\gamma}$. A difficulty encountered in our data analysis has been the presence of an excess of residual electronic systematics in the spectral analyser we used, which we have attempted to reduce using a Principal Component Analysis Method. This has helped reduce the level of systematics but it has still caused problems in degrading our sensitivity; nevertheless, the problem should be resolved with improved equipment and experimental protocols in upcoming experiments in RADES.  

Additional RADES R\&D initiatives are under progress, including testing taller HTS-coated cavities as well as mechanical \cite{Golm:2023iwe}, ferroelectric and ferromagnetic tuning \cite{10078243, JMGBThesis} that should allow a search over a broader frequency range. 

On the very long term, RADES envisages a data-taking in the magnet of the babyIAXO haloscope, with striking sensitivity in the 1-2 $\mu$eV mass range \cite{Ahyoune:2023gfw}.

\section*{Acknowledgements}

This project has received funding from the European Union’s Horizon 2020
Research and Innovation programme under Grant Agreement No 730871
(ARIES-TNA). This work has also received funding through the European
Research Council under grant ERC-2018-StG-802836 (AxScale). IGI acknowledges funding through the European Research Council under grant ERC-2017-AdG-788781 (IAXO+).

This work has been partially funded by MCIN/AEI/10.13039/501100011033/ and by "ERDF A way of making Europe", under grant PID2019-108122GB-C33, by PID2019-108122GB-C32, and by Maria de Maeztu program CEX2019-000918-M. JMGB thanks the grant FPI BES-2017-079787, funded by MCIN/AEI/10.13039/501100011033 and by "ESF Investing in your future". 

Generalitat Valenciana has also funded this work under the project ASFAE/2022/013.

The authors acknowledge the support and samples provided
by THEVA.

ICMAB co-authors acknowledges funds from  PID2021-127297OB-C21 and CEX2019-000917-S, FCC-GOV-CC-0208 (KE4947/ATS) and PRTR-C17I1 from MICIIN-NGEU-Generalitat de Catalunya.

We thank Giuseppe Ruoso for the loan of equipment and the CERN teams at SM18 and in the CERN Central Cryogenic Laboratory for the support, particularly: Gerard Willering, Franco Julio Mangiarotti, Jerome Feuvrier, Marta Bajko, Patrick Viret, Guillaume Pichon, Arnaud Devred, Stephan Russenschuck,  Andrzej Siemko, Joanna Liberadzka-Porret and Torsten Koettig.

\bibliographystyle{JHEP_improved}
\bibliography{biblio}

\providecommand{\href}[2]{#2}\begingroup\raggedright\begin{thebibliography}{10}

\bibitem{Peccei:1977hh}
R.~D. Peccei and H.~R. Quinn,
  \href{http://dx.doi.org/10.1103/PhysRevLett.38.1440}{{\it {CP Conservation in
  the presence of pseudoparticles}}, } {\em Phys. Rev. Lett.} {\bf 38} (1977)
  1440--1443.

\bibitem{Wilczek:1977pj}
F.~Wilczek, \href{http://dx.doi.org/10.1103/PhysRevLett.40.279}{{\it {Problem
  of Strong $P$ and $T$ Invariance in the Presence of Instantons}}, } {\em
  Phys. Rev. Lett.} {\bf 40} (1978) 279--282.

\bibitem{Weinberg:1977ma}
S.~Weinberg, \href{http://dx.doi.org/10.1103/PhysRevLett.40.223}{{\it {A New
  Light Boson?}}, } {\em Phys. Rev. Lett.} {\bf 40} (1978) 223--226.

\bibitem{Ipser:1983mw}
J.~Ipser and P.~Sikivie,
  \href{http://dx.doi.org/10.1103/PhysRevLett.50.925}{{\it {Are Galactic Halos
  Made of Axions?}}, } {\em Phys. Rev. Lett.} {\bf 50} (1983) 925.

\bibitem{Turner:1983sj}
M.~S. Turner, F.~Wilczek, and A.~Zee,
  \href{http://dx.doi.org/10.1016/0370-2693(83)91229-7}{{\it {Formation of
  Structure in an Axion Dominated Universe}}, } {\em Phys. Lett. B} {\bf 125}
  (1983) 35. [Erratum: Phys.Lett.B 125, 519 (1983)].

\bibitem{Preskill:1982cy}
J.~Preskill, M.~B. Wise, and F.~Wilczek,
  \href{http://dx.doi.org/10.1016/0370-2693(83)90637-8}{{\it {Cosmology of the
  Invisible Axion}}, } {\em Phys. Lett. B} {\bf 120} (1983) 127--132.

\bibitem{Abbott:1982af}
L.~F. Abbott and P.~Sikivie,
  \href{http://dx.doi.org/10.1016/0370-2693(83)90638-X}{{\it {A Cosmological
  Bound on the Invisible Axion}}, } {\em Phys. Lett. B} {\bf 120} (1983)
  133--136.

\bibitem{Dine:1982ah}
M.~Dine and W.~Fischler,
  \href{http://dx.doi.org/10.1016/0370-2693(83)90639-1}{{\it {The Not So
  Harmless Axion}}, } {\em Phys. Lett. B} {\bf 120} (1983) 137--141.

\bibitem{Klaer:2017ond}
V.~B.~. Klaer and G.~D. Moore,
  \href{http://dx.doi.org/10.1088/1475-7516/2017/11/049}{{\it {The dark-matter
  axion mass}}, } {\em JCAP} {\bf 11} (2017) 049,
  [\href{http://arxiv.org/abs/1708.07521}{{\tt 1708.07521}}].

\bibitem{Saikawa:2024bta}
K.~Saikawa, J.~Redondo, A.~Vaquero, and M.~Kaltschmidt, {\it {Spectrum of
  global string networks and the axion dark matter mass}},
  \href{http://arxiv.org/abs/2401.17253}{{\tt 2401.17253}}.

\bibitem{ohare}
C.~O'HARE, ``Axion limits.'' \url{https://github.com/cajohare/AxionLimits/}.

\bibitem{Pirmakoff:1951pj}
H.~Primakoff, \href{http://dx.doi.org/10.1103/PhysRev.81.899}{{\it
  {Photoproduction of neutral mesons in nuclear electric fields and the mean
  life of the neutral meson}}, } {\em Phys. Rev.} {\bf 81} (1951) 899.

\bibitem{Sikivie:1983ip}
P.~Sikivie, \href{http://dx.doi.org/10.1103/PhysRevLett.51.1415}{{\it
  {Experimental Tests of the Invisible Axion}}, } {\em Phys. Rev. Lett.} {\bf
  51} (1983) 1415--1417. [Erratum: Phys.Rev.Lett. 52, 695 (1984)].

\bibitem{ADMX:2020ote}
{\bf ADMX Collaboration}, R.~Khatiwada et~al.,
  \href{http://dx.doi.org/10.1063/5.0037857}{{\it {Axion Dark Matter
  Experiment: Detailed design~and operations}}, } {\em Rev. Sci. Instrum.} {\bf
  92} (2021), no.~12 124502, [\href{http://arxiv.org/abs/2010.00169}{{\tt
  2010.00169}}].

\bibitem{Irastorza:2018dyq}
I.~G. Irastorza and J.~Redondo,
  \href{http://dx.doi.org/10.1016/j.ppnp.2018.05.003}{{\it {New experimental
  approaches in the search for axion-like particles}}, } {\em Prog. Part. Nucl.
  Phys.} {\bf 102} (2018) 89--159, [\href{http://arxiv.org/abs/1801.08127}{{\tt
  1801.08127}}].

\bibitem{ADMX:2021nhd}
{\bf ADMX Collaboration}, C.~Bartram et~al.,
  \href{http://dx.doi.org/10.1103/PhysRevLett.127.261803}{{\it {Search for
  Invisible Axion Dark Matter in the 3.3\textendash{}4.2\,\,\ensuremath{\mu}eV
  Mass Range}}, } {\em Phys. Rev. Lett.} {\bf 127} (2021), no.~26 261803,
  [\href{http://arxiv.org/abs/2110.06096}{{\tt 2110.06096}}].

\bibitem{Adair:2022rtw}
C.~M. Adair et~al., \href{http://dx.doi.org/10.1038/s41467-022-33913-6}{{\it
  {Search for Dark Matter Axions with CAST-CAPP}}, } {\em Nature Commun.} {\bf
  13} (2022), no.~1 6180, [\href{http://arxiv.org/abs/2211.02902}{{\tt
  2211.02902}}].

\bibitem{Backes:2020ajv}
{\bf HAYSTAC Collaboration}, K.~M. Backes et~al.,
  \href{http://dx.doi.org/10.1038/s41586-021-03226-7}{{\it {A quantum-enhanced
  search for dark matter axions}}, } {\em Nature Phys.} {\bf 590} (2021),
  no.~7845 238--242, [\href{http://arxiv.org/abs/2008.01853}{{\tt
  2008.01853}}].

\bibitem{Alesini:2022lnp}
D.~Alesini et~al., \href{http://dx.doi.org/10.1103/PhysRevD.106.052007}{{\it
  {Search for Galactic axions with a high-Q dielectric cavity}}, } {\em Phys.
  Rev. D} {\bf 106} (2022), no.~5 052007,
  [\href{http://arxiv.org/abs/2208.12670}{{\tt 2208.12670}}].

\bibitem{Quiskamp:2022pks}
A.~P. Quiskamp, B.~T. McAllister, P.~Altin, E.~N. Ivanov, M.~Goryachev, et~al.,
  \href{http://dx.doi.org/10.1126/sciadv.abq3765}{{\it {Direct search for dark
  matter axions excluding ALP cogenesis in the 63- to 67-\ensuremath{\mu}eV
  range with the ORGAN experiment}}, } {\em Sci. Adv.} {\bf 8} (2022), no.~27
  abq3765, [\href{http://arxiv.org/abs/2203.12152}{{\tt 2203.12152}}].

\bibitem{Garcia-Barcelo:2023iri}
J.~M. Garc\'\i{}a-Barcel\'o, A.~D\'\i{}az-Morcillo, and B.~Gimeno,
  \href{http://dx.doi.org/10.1007/JHEP11(2023)159}{{\it {Enhancing resonant
  circular-section haloscopes for dark matter axion detection: approaches and
  limitations in volume expansion}}, } {\em JHEP} {\bf 11} (2023) 159,
  [\href{http://arxiv.org/abs/2309.13199}{{\tt 2309.13199}}].

\bibitem{Garcia-Barcelo:2023wrw}
J.~M. Garc\'\i{}a-Barcel\'o et~al.,
  \href{http://dx.doi.org/10.1007/JHEP08(2023)098}{{\it {Methods and
  restrictions to increase the volume of resonant rectangular-section
  haloscopes for detecting dark matter axions}}, } {\em JHEP} {\bf 08} (2023)
  098, [\href{http://arxiv.org/abs/2302.10569}{{\tt 2302.10569}}].

\bibitem{Ivanov:2022hlb}
{\bf MADMAX Collaboration}, A.~Ivanov, C.~Lee, X.~Li, O.~Reimann, and D.~Strom,
  \href{http://dx.doi.org/10.22323/1.414.0109}{{\it {MADMAX - A Novel
  dielectric haloscope detector for post-inflationary axion dark matter
  searches}}, } {\em PoS} {\bf ICHEP2022} (11, 2022) 109.

\bibitem{ALPHA:2022rxj}
{\bf ALPHA Collaboration}, A.~J. Millar et~al.,
  \href{http://dx.doi.org/10.1103/PhysRevD.107.055013}{{\it {Searching for dark
  matter with plasma haloscopes}}, } {\em Phys. Rev. D} {\bf 107} (2023), no.~5
  055013, [\href{http://arxiv.org/abs/2210.00017}{{\tt 2210.00017}}].

\bibitem{CAST:2020rlf}
{\bf CAST Collaboration}, A.~A. Melc\'on et~al.,
  \href{http://dx.doi.org/10.1007/JHEP10(2021)075}{{\it {First results of the
  CAST-RADES haloscope search for axions at 34.67 $\mu$eV}}, } {\em JHEP} {\bf
  21} (2020) 075, [\href{http://arxiv.org/abs/2104.13798}{{\tt 2104.13798}}].

\bibitem{Magnet}
F.~Savary, G.~Apollinari, B.~Auchmann, E.~Barzi, G.~Chlachidze, et~al.,
  \href{http://dx.doi.org/10.1109/TASC.2015.2395381}{{\it {Design, Assembly,
  and Test of the CERN 2-m Long 11 T Dipole in Single Coil Configuration}}, }
  {\em IEEE Transactions on Applied Superconductivity} {\bf 25} (2015), no.~3
  1--5.

\bibitem{capptalk}
W.~Chung, ``Capp-max.'' 18th PATRAS workshop, 2023.

\bibitem{Golm:2021ooj}
J.~Golm et~al., \href{http://dx.doi.org/10.1109/TASC.2022.3147741}{{\it {Thin
  Film (High Temperature) Superconducting Radiofrequency Cavities for the
  Search of Axion Dark Matter}}, } {\em IEEE Trans. Appl. Supercond.} {\bf 32}
  (2022), no.~4 1500605, [\href{http://arxiv.org/abs/2110.01296}{{\tt
  2110.01296}}].

\bibitem{Alesini:2019ajt}
D.~Alesini et~al., \href{http://dx.doi.org/10.1103/PhysRevD.99.101101}{{\it
  {Galactic axions search with a superconducting resonant cavity}}, } {\em
  Phys. Rev. D} {\bf 99} (2019), no.~10 101101,
  [\href{http://arxiv.org/abs/1903.06547}{{\tt 1903.06547}}].

\bibitem{Ahn:2021fgb}
D.~Ahn, O.~Kwon, W.~Chung, W.~Jang, D.~Lee, et~al.,
  \href{http://dx.doi.org/10.1103/PhysRevApplied.17.L061005}{{\it {Biaxially
  Textured YBa2Cu3O7\ensuremath{-}x Microwave Cavity in a High Magnetic Field
  for a Dark-Matter Axion Search}}, } {\em Phys. Rev. Applied} {\bf 17} (2022),
  no.~6 L061005, [\href{http://arxiv.org/abs/2103.14515}{{\tt 2103.14515}}].

\bibitem{AlvarezMelcon:2020vee}
A.~\'Alvarez~Melc\'on et~al.,
  \href{http://dx.doi.org/10.1007/JHEP07(2020)084}{{\it {Scalable haloscopes
  for axion dark matter detection in the 30$\mu$eV range with RADES}}, } {\em
  JHEP} {\bf 07} (2020) 084, [\href{http://arxiv.org/abs/2002.07639}{{\tt
  2002.07639}}].

\bibitem{PRUSSEIT2005866}
W.~Prusseit, R.~Nemetschek, C.~Hoffmann, G.~Sigl, A.~Lümkemann, et~al.,
  \href{https://www.sciencedirect.com/science/article/pii/S092145340500376X}{{\it
  {ISD process development for coated conductors}}, } {\em Physica C:
  Superconductivity and its Applications} {\bf 426-431} (2005) 866--871.
  Proceedings of the 17th International Symposium on Superconductivity (ISS
  2004).

\bibitem{ICMAB}
``{ICMAB} webpage.'' \url{https://icmab.es/}.
\newblock Accessed: 2024-03-06.

\bibitem{Romanov:2020epk}
A.~Romanov, P.~Krkoti\'c, G.~Telles, J.~O'Callaghan, M.~Pont, et~al.,
  \href{http://dx.doi.org/10.1038/s41598-020-69004-z}{{\it {High frequency
  response of thick REBCO coated conductors in the framework of the FCC
  study}}, } {\em Sci. Rep.} {\bf 10} (2020), no.~1 12325.

\bibitem{Telles_2023}
G.~T. Telles, A.~Romanov, S.~Calatroni, X.~Granados, T.~Puig, et~al.,
  \href{https://dx.doi.org/10.1088/1361-6668/ac97c9}{{\it {Field quality and
  surface resistance studies of a superconducting REBa2Cu3O—Cu hybrid coating
  for the FCC beam screen}}, } {\em Superconductor Science and Technology} {\bf
  36} (feb, 2023) 045001.

\bibitem{RomanovThesis}
A.~Romanov, {\em {Superconducting Coated Conductors for Proton Beam Screens in
  High-Energy Particle Accelerators}}.
\newblock PhD thesis, Universitat Autònoma de Barcelona, 2022.

\bibitem{Otten_2016}
S.~Otten, A.~Kario, A.~Kling, and W.~Goldacker,
  \href{https://dx.doi.org/10.1088/0953-2048/29/12/125003}{{\it Bending
  properties of different rebco coated conductor tapes and roebel cables at t =
  77 k}, } {\em Superconductor Science and Technology} {\bf 29} (oct, 2016)
  125003.

\bibitem{EKIN}
J.~Ekin, {\em Experimental techniques for low-temperature measurements:
  cryostat design, material properties and superconductor critical-current
  testing}.
\newblock Oxford university press, 2006.

\bibitem{CST}
``Cst studio suite: Electromagnetic field simulation software.''
  \url{https://www.solidworks.com/de/media/cst-studio-suite-electromagnetic-field-simulation-software}.
\newblock Accessed: 2023-07-06.

\bibitem{lownoisefactory}
``{Datasheet 6-20 GHz Cryogenic Low Noise Amplifier}.''
  \url{https://lownoisefactory.com/wp-content/uploads/2022/03/lnf-lnc6_20c.pdf}.
\newblock Accessed: 2023-03-07.

\bibitem{TTI}
``{Radiofrequency and Antenna solutions}.'' \url{https://www.ttinorte.es/}.
\newblock Accessed: 2023-03-07.

\bibitem{SACThesis}
S.~Arguedas~Cuendis, {\em {Dark Matter axion search using novel RF resonant
  cavity geometries in the CAST experiment}}.
\newblock PhD thesis, University of Vienna, 2021.

\bibitem{software_products}
``Fluid property packages.'' \url{https://htess.com/software-products/}.
\newblock Accessed: 2023-04-17.

\bibitem{pozar2011microwave}
D.~M. Pozar, {\em Microwave engineering}.
\newblock John wiley \& sons, 2011.

\bibitem{datasheet_cables}
{Rf microwave electronic components shop: Datasheet of Semirigid PTFE Microwave
  Coaxial Cable - RG402-Cu }.
\newblock Accessed Jan. 10, 2023.

\bibitem{Y-method}
K.~Technologies, ``Application {N}ote 5952-3706{E} (2019).''
  \url{https://www.keysight.com/ch/de/assets/7018-06829/application-notes/5952-3706.pdf}.

\bibitem{Brubaker:2017rna}
B.~M. Brubaker et~al.,
  \href{https://link.aps.org/doi/10.1103/PhysRevD.96.123008}{{\it Haystac axion
  search analysis procedure}, } {\em Phys. Rev. D} {\bf 96} (Dec, 2017) 123008.

\bibitem{PCA}
W.~J. Krzanowski, \href{https://doi.org/10.2307/2982678}{{\it {A User’s Guide
  to Principal Components}}, } {\em Journal of the Royal Statistical Society
  Series A: Statistics in Society} {\bf 155} (12, 2018) 173--174,
  [\href{http://arxiv.org/abs/https://academic.oup.com/jrsssa/article-pdf/155/1/173/49762607/jrsssa\_155\_1\_173.pdf}{{\tt
  https://academic.oup.com/jrsssa/article-pdf/155/1/173/49762607/jrsssa\_155\_1\_173.pdf}}].

\bibitem{savitzky64}
A.~Savitzky and M.~J.~E. Golay, {\it Smoothing and differentiation of data by
  simplified least squares procedures},  {\em Anal. Chem.} {\bf 36} (1964)
  1627--1639.

\bibitem{Turner:1990qx}
M.~S. Turner, \href{https://link.aps.org/doi/10.1103/PhysRevD.42.3572}{{\it
  Periodic signatures for the detection of cosmic axions}, } {\em Phys. Rev. D}
  {\bf 42} (Nov, 1990) 3572--3575.

\bibitem{Read:2014qva}
J.~I. Read, \href{http://dx.doi.org/10.1088/0954-3899/41/6/063101}{{\it {The
  Local Dark Matter Density}}, } {\em J. Phys. G} {\bf 41} (2014) 063101,
  [\href{http://arxiv.org/abs/1404.1938}{{\tt 1404.1938}}].

\bibitem{Alesini:2020vny}
D.~Alesini et~al., {\it {Search for Invisible Axion Dark Matter of mass
  m$_a=43~\mu$eV with the QUAX--$a\gamma$ Experiment}},
  \href{http://arxiv.org/abs/2012.09498}{{\tt 2012.09498}}.

\bibitem{Boutan:2018uoc}
C.~Boutan et~al., \href{http://dx.doi.org/10.1103/PhysRevLett.121.261302}{{\it
  Piezoelectrically tuned multimode cavity search for axion dark matter}, }
  {\em Phys. Rev. Lett.} {\bf 121} (12, 2018).

\bibitem{DePanfilis:1987dk}
S.~De~Panfilis et~al., \href{http://dx.doi.org/10.1103/PhysRevLett.59.839}{{\it
  {Limits on the Abundance and Coupling of Cosmic Axions at 4.5-Microev
  \ensuremath{<} m(a) \ensuremath{<} 5.0-Microev}}, } {\em Phys. Rev. Lett.}
  {\bf 59} (1987) 839.

\bibitem{Hagmann:1990tj}
C.~Hagmann et~al., \href{http://dx.doi.org/10.1103/PhysRevD.42.1297}{{\it
  {Results from a search for cosmic axions}}, } {\em Phys. Rev. D} {\bf 42}
  (1990) 1297--1300.

\bibitem{Braine:2019fqb}
{\bf ADMX Collaboration}, T.~Braine et~al.,
  \href{http://dx.doi.org/10.1103/PhysRevLett.124.101303}{{\it {Extended Search
  for the Invisible Axion with the Axion Dark Matter Experiment}}, } {\em Phys.
  Rev. Lett.} {\bf 124} (2020), no.~10 101303,
  [\href{http://arxiv.org/abs/1910.08638}{{\tt 1910.08638}}].

\bibitem{Du:2018uak}
{\bf ADMX Collaboration}, N.~Du et~al.,
  \href{http://dx.doi.org/10.1103/PhysRevLett.120.151301}{{\it {A Search for
  Invisible Axion Dark Matter with the Axion Dark Matter Experiment}}, } {\em
  Phys. Rev. Lett.} {\bf 120} (2018), no.~15 151301,
  [\href{http://arxiv.org/abs/1804.05750}{{\tt 1804.05750}}].

\bibitem{Lee:2020cfj}
S.~Lee et~al., \href{http://dx.doi.org/10.1103/PhysRevLett.124.101802}{{\it
  {Axion Dark Matter Search around 6.7 $\mu$eV}}, } {\em Phys. Rev. Lett.} {\bf
  124} (2020), no.~10 101802, [\href{http://arxiv.org/abs/2001.05102}{{\tt
  2001.05102}}].

\bibitem{Jeong:2020cwz}
J.~Jeong et~al., \href{http://dx.doi.org/10.1103/PhysRevLett.125.221302}{{\it
  {Search for Invisible Axion Dark Matter with a Multiple-Cell Haloscope}}, }
  {\em Phys. Rev. Lett.} {\bf 125} (2020), no.~22 221302,
  [\href{http://arxiv.org/abs/2008.10141}{{\tt 2008.10141}}].

\bibitem{Zhong:2018rsr}
{\bf HAYSTAC Collaboration}, L.~Zhong et~al.,
  \href{http://dx.doi.org/10.1103/PhysRevD.97.092001}{{\it {Results from phase
  1 of the HAYSTAC microwave cavity axion experiment}}, } {\em Phys. Rev. D}
  {\bf 97} (2018), no.~9 092001, [\href{http://arxiv.org/abs/1803.03690}{{\tt
  1803.03690}}].

\bibitem{McAllister:2017lkb}
B.~T. McAllister et~al.,
  \href{http://dx.doi.org/10.1016/j.dark.2017.09.010}{{\it {The ORGAN
  Experiment: An axion haloscope above 15 GHz}}, } {\em Phys. Dark Univ.} {\bf
  18} (2017) 67--72, [\href{http://arxiv.org/abs/1706.00209}{{\tt
  1706.00209}}].

\bibitem{Anastassopoulos:2017ftl}
{\bf CAST Collaboration}, V.~Anastassopoulos et~al.,
  \href{http://dx.doi.org/10.1038/nphys4109}{{\it {New CAST Limit on the
  Axion-Photon Interaction}}, } {\em Nature Phys.} {\bf 13} (2017) 584--590,
  [\href{http://arxiv.org/abs/1705.02290}{{\tt 1705.02290}}].

\bibitem{PBCtalk_JGolm}
J.~Golm, ``\href{https://indico.cern.ch/event/1137276}{HTS RF cavities for
  axion searches and RADES update}.'' Physics Beyond Colliders Annual Workshop,
  2022.

\bibitem{Golm:2023iwe}
J.~Golm, J.~M. Garc\'\i{}a-Barcel\'o, S.~Calatroni, W.~Wuensch, and
  B.~D\"obrich, {\it {Mechanical tuning of a rectangular axion haloscope
  operating around 8.4 GHz}},  \href{http://arxiv.org/abs/2312.13109}{{\tt
  2312.13109}}.

\bibitem{10078243}
J.~M. García-Barceló et~al.,
  \href{http://dx.doi.org/10.1109/ACCESS.2023.3260783}{{\it On the development
  of new tuning and inter-coupling techniques using ferroelectric materials in
  the detection of dark matter axions}, } {\em IEEE Access} {\bf 11} (2023)
  30360--30372.

\bibitem{JMGBThesis}
J.~M. García-Barceló, {\em {Development of resonant cavity-based microwave
  filters for axion detection}}.
\newblock PhD thesis, Technical University of Cartagena, 2023.

\bibitem{Ahyoune:2023gfw}
S.~Ahyoune et~al., \href{http://dx.doi.org/10.1002/andp.202300326}{{\it {A
  Proposal for a Low-Frequency Axion Search in the 1\textendash{}2
  \ensuremath{\mu} eV Range and Below with the BabyIAXO Magnet}}, } {\em
  Annalen Phys.} {\bf 535} (2023), no.~12 2300326,
  [\href{http://arxiv.org/abs/2306.17243}{{\tt 2306.17243}}].

\end{thebibliography}\endgroup

\end{document}